\documentclass[11pt]{article}

\usepackage{graphicx}
\usepackage{multicol}
\usepackage{color}
\usepackage{epsfig,latexsym}

\definecolor{verdon}{cmyk}{1,0.5,1,0}
\definecolor{blue}{cmyk}{0.8,0.8,0,0.}
\definecolor{red}{cmyk}{0.2,1,1,0.0}
\newcommand{\cita}[1]{{\color{blue} \cite{#1}}}

\def\lapprox{\mathrel{\mathop  {\hbox{\lower0.5ex\hbox{$\sim$}
\kern-1.1em\lower-0.7ex\hbox{$<$}}}}}
\def\gapprox{\mathrel{\mathop  {\hbox{\lower0.5ex\hbox{$\sim$}
\kern-1.1em\lower-0.7ex\hbox{$>$}}}}}

\begin{document}

\title{\color{verdon} Linear Solar Models}
\author{F.L. Villante$^{1}$ and B. Ricci$^2$
\vspace{0.5 cm}\\
{\small\em $^{1}$Universit\`a dell'Aquila and INFN - LNGS, L'Aquila - Italy} \\
{\small\em $^{2}$Universit\`a di Ferrara and INFN, Ferrara - Italy}}

\date{}

\maketitle

\def\abstractname{\color{red}\bf Abstract}
\begin{abstract}
{\footnotesize
We present a new approach to study the properties of the sun.
We consider small variations of the physical and chemical properties of the sun 
with respect to Standard Solar Model predictions and we linearize the structure equations  to relate them
to the properties of the solar plasma.  By assuming that the (variation of) the present solar composition 
can be estimated from the (variation of) the nuclear reaction rates and elemental diffusion efficiency 
in the present sun, we obtain a linear system of ordinary differential equations which can 
be used to calculate the response of the sun to an arbitrary modification
of the input parameters (opacity, cross sections, etc.).
This new approach is intended to be a complement to the traditional 
methods for solar model calculation and allows to investigate in a more efficient and 
transparent way the role of parameters and assumptions
 in solar model construction. 
We verify that these Linear Solar Models recover the predictions
of the traditional solar models with an high level of accuracy.}
\end{abstract}

\newpage

\section{Introduction}

In the last three decades, there was an enormous progress in our understanding of the sun.
The predictions of the Standard Solar Model (SSM), which is the fundamental theoretical tool
to investigate the solar interior, have been tested by solar neutrino experiments and
by helioseismology.

The deficit of the observed solar neutrino fluxes, reported initially by Homestake \cita{Homestake}
 and then confirmed by GALLEX \cita{Gallex} and SAGE \cita{SAGE} (which subsequently merged
into GNO \cita{GNO}), Kamiokande \cita{Kamiokande} and
Super-Kamiokande \cita{SK}, generated the so-called ``solar neutrino
problem''  which stimulated a deep investigation of the solar
structure (see e.g. \cita{SunReviews}). The problem was solved in
2002 when the
SNO experiment \cita{SNO}
obtained a direct evidence for flavour oscillations
of solar neutrinos and, moreover, confirmed the SSM prediction of the $^{8}{\rm B}$ neutrino flux with
an accuracy which, according to the latest data \cita{SNOLatest}, is equal to about $\sim 6 \%$ (see e.g. \cita{Serenelli09})\footnote{
For the sake of precision, the first model-independent evidence for solar neutrino oscillations and
the first determination of the $^{8}{\rm B}$ solar neutrino flux has been obtained in 2001 (see e.g. \cita{elio}) 
by comparing the SNO charged-current result \cita{SNO-CC} with the SK data with
the method proposed by \cita{Villante98}. The year 2002 is, however, recognized as the 'annus mirabilis' \cita{elio2}
for the solar neutrino physics. During 2002, in fact, the SNO neutral-current measurement \cita{SNO} and
the first KamLAND results \cita{KamLAND1} were released and, moreover, Raymond Davis, Jr. and 
Masatoshi Koshiba were awarded with the Nobel Prize.}.

At the same time, helioseismic observations have allowed to determine precisely several important
properties of the sun, such as the depth of the convective envelope which is known at the $\sim 0.2\%$ level,
the surface helium abundance which is obtained at the $\sim 1.5 \%$ level
and the sound speed profile which is determined with an accuracy equal to $\sim 0.1 \%$
in a large part of the sun (see e.g. \cita{HelioReview,Basu:2007fp} and references therein).
 As a results of these observations, the solar structure is
now very well constrained, so that the sun can be used as a solid benchmark for stellar evolution
and as a ``laboratory'' for fundamental physics (see e.g. \cita{SunLab}).

The future times could be even more interesting. KamLAND reactor (anti)neutrino
experiment \cita{KamLAND} has confirmed the flavour oscillations hypothesis and has refined the determination of neutrino
parameters. We now reliably know the solar neutrino oscillation probability and we
 can go back to the original program of solar neutrino studies, i.e. to probe nuclear reactions in the solar core.
The present and future solar neutrino experiments, such as Borexino \cita{Borexino} and SNO+ \cita{SNO+}, have the potential to
provide the first direct measurements of the CNO and pep neutrinos, thus probing
dominant and sub-dominant energy generation mechanisms in the sun.

At the same time, a new solar problem has emerged. Recent determinations of the photospheric
heavy element abundances \cita{as05,as09} indicate that the sun metallicity is lower than previously
assumed \cita{gs98}. Solar models that incorporate these lower abundances are no more able to
reproduce the helioseismic results. As an example, the sound speed predicted by SSMs at the bottom
of the convective envelope disagrees at the $\sim 1\%$ level with the value inferred by helioseismic data (see e.g.\cita{BBPS05}).
Detailed studies have been done to resolve this
controversy (see e.g.\cita{ Basu:2007fp,as09}), but a definitive solution of the ``solar composition problem''
still has to be obtained.

In this framework, it is important to analyze of the role of physical inputs and of
the standard assumptions for SSM calculations. This task is not always possible in simple and clear terms,
since SSM construction relies on (time-consuming) numerical integration of a non linear system
of partial differential equations. Several input parameters are
necessary to fully describe the property of the solar plasma and some of them are not single numbers
but complicated functions (like e.g. the opacity of the solar interior) which, in principle,
can be modified in a non trivial way. Moreover, any modification of the sun produces a variety of correlated
effects
which have to be taken into account all together if we want to correctly extract information from
the comparison of theoretical predictions with observational data.

In order to overcome these difficulties, we provide a tool which is at same time simple and accurate enough to
describe the effects
of a generic (small) modification of the physical inputs.
The starting point is the fact that, despite the
present disagreement with helioseismic data, the SSM is a rather good approximation of the real sun.
We can thus assume small variations of the physical and chemical properties of the sun with respect to the SSM predictions
and use a linear theory to relate them to
the properties of the solar plasma.
With the additional assumption that the (variation of) the present solar composition can be estimated from the
(variation of) the nuclear reaction rate and elemental diffusion efficiency in the present sun,
 we obtain a linear system of ordinary differential
equations, which can be easily integrated.

We believe that the proposed approach can be useful in several respects.
First, the construction of these Linear Solar Models (LSM) can complement the traditional methods for 
solar model calculations,
allowing to investigate in a more efficient and transparent way the role of parameters and assumptions.
 In a separated paper, we will use LSMs to discuss in general terms the role of opacity and metals
 in the solar interior \cita{noiNext}. Moreover,  it can help to introduce new effects and to understand their
 relevance, prior to the implementation into the more complicated SSM machinery.
Finally, this simplified approach could open the field of solar physics
beyond the small community of the SSM's builders
with the result of making the sun a more accessible ``laboratory''.
Clearly, LSMs have not to be intended as an alternative to SSMs, which remain the
fundamental theoretical tool to compare with observations.

 The plan of the paper is the following. In the next section we briefly review SSM calculations.
In sect.\ref{Linear} we expand to linear order the structure equations of the present sun.
In sect.\ref{Properties} we calculate the properties of the sun and of the solar plasma that
are necessary to define LSMs. In sect.\ref{Boundary} we give the integration conditions for 
the linearized equations.
In sect.\ref{chemical}, we discuss how to estimate the chemical composition of the sun. In sect.\ref{Fullset}
we give the equations that define LSMs in their final form. Finally, in sect.\ref{Comparison}
we compare the results of LSMs with those obtained by using the standard method for solar model 
calculations, showing the validity of our simplified approach.

\section{The standard solar model}
\label{SSM}

In the assumption of spherical symmetry, the behavior of the pressure ($P$), density ($\rho$), temperature ($T$), luminosity ($l$)
and mass ($m$)  in the sun is described by the structure equations \cita{reviews}:
\begin{eqnarray}
\nonumber
    \frac{\partial m}{\partial r} &=& 4\pi r^{2} \rho \label{eq1}\\
\nonumber
    \frac{\partial P}{\partial r} &=& - \frac{G_{\rm N} m} {r^2} \rho \label{eq2}\\
\nonumber
    \frac{\partial l}{\partial r} &=& 4\pi r^{2} \rho \; \epsilon(\rho,T,X_{i}) \label{eq3}\\
\nonumber
    \frac{\partial T}{\partial r} &=& -\frac{G_{\rm N} m T \rho}{r^2 P} \nabla \label{eq4}\\
    P &=& P(\rho,T,X_{i}) \label{eq5}
\end{eqnarray}
where $r$ indicates the distance from the center, $\epsilon(\rho,T,X_{i})$ is
the energy produced per unit time and mass\footnote{We neglect, here and in the following, the
contribution to energy balance due to the heat released by the various shells of the sun in their
thermodynamical transformations. This contribution, which depends on the time derivative of pressure and
temperature, is sub-dominant as can be understood from the fact that the sun evolves on time
scales much larger than the Kelvin-Helmotz time.},
the function $P= P(\rho,T,X_{i})$ describes the
equation of state (EOS) of solar matter and
$X_{i}(r)$ are the mass abundances of the various chemical elements
inside the sun.
In the equation which describes the energy transport, the
temperature gradient $\nabla$ is defined as $\nabla \equiv \partial \ln T/ \partial \ln P$.
If the energy transfer is due to radiative processes (i.e. at the center of the sun) one has:
\begin{equation}
\nabla = \nabla_{\rm rad}
\end{equation}
where the radiative gradient $\nabla_{\rm rad}$ is given by:
\begin{equation}
\nabla_{\rm rad}=\frac{3}{16\pi ac \, G_{\rm N}}
\frac{\kappa(\rho,T,X_{i}) \,l \, P}{m \, T^4}
\end{equation}
and  $\kappa(\rho,T,X_{i})$ is the opacity of solar matter.
In the presence of convective motions
(i.e. in the outer layer of the sun),
the value of  $\nabla$ has to be calculated by taking
into account the convective energy transport which generally provides the dominant contribution.
In a large part of the sun convective envelope, one has
\begin{equation}
\nabla\simeq\nabla_{\rm ad}\simeq 0.4~.
\end{equation}
This is expected when convection is very efficient
and a negligible excess of $\nabla$ over the adiabatic value $\nabla_{\rm ad}$
is sufficient to transport the whole luminosity. In the outermost layers of the sun, the situation
is more complicated. The precise description of convection in this regime is still an
unsolved problem. One uses a phenomenological model which
predicts the efficiency of convection as a function of the "mixing length parameter"
$\alpha$, which is related to the distances over which a moving unit of gas can be identified
before it mixes appreciably (see e.g. \cite{reviews}).
The transition between the internal radiative core and the external convective
envelope occurs at the radius $R_{\rm b}$ where:
\begin{equation}
\nabla_{\rm rad}(R_{\rm b}) = 
\nabla_{\rm ad} \; ,
\end{equation}
as it is prescribed by the Schwarzschild criterion which states that convective motions occur
 in the region of the star where $\nabla_{\rm rad}(r) > \nabla_{\rm ad}$.

The chemical composition of the present sun is not known
but has to be calculated by coupling eqs.~(\ref{eq5}), which describe
the mechanical and thermal structure, with the equations that describe the
chemical evolution of the sun (see e.g. \cita{reviews} for details).
It is generally assumed that the sun was born chemically homogeneous
and has modified its chemical composition due to nuclear reactions and elemental diffusion.
One assumes that {\em relative} heavy element abundances in the photosphere
have not been modified during the evolution and uses the observationally
determined photospheric composition (see e.g.\cita{as05,as09}) to fix the initial
heavy element admixture.  We remark that the chemical abundances are uniform in the 
convective region due to the very efficient mixing induced by convective motions. 
The values $X_{i, \rm b}$, evaluated at the bottom of the convective region, 
are, thus, representative for the solar surface composition.

If a complete information about the initial composition was
available and if a complete theory of convection was known, then there
would be no free parameter. In practice, one has three free parameters, namely the initial metal 
abundance $Z_{\rm ini}$, the initial helium abundance $Y_{\rm ini}$ and the mixing length parameter $\alpha$.
These are tuned in order to reproduce, at the solar age $t_{\odot}=4.57 \; {\rm Gyr}$ \cita{age},
the observed solar luminosity $L_{\odot}= 3.8418 \cdot 10^{33}$erg/s \cita{lum}, 
the observed solar radius $R_{\odot}= 6.9598 \cdot 10^{10}$cm
\cita{radius} and the surface metal-to-hydrogen ratio $(Z/X)_{\rm b}$. 
In this paper we use the AS05 composition \cita{as05} which corresponds to 
$(Z/X)_{\rm b}=0.0165$.

A standard solar model (SSM), according to the definition of \cita{boh},
is a solution of the above problem which reproduces,
within uncertainties, the observed
properties of the sun, by adopting physical and chemical inputs chosen
within their range of uncertainties.
 In this paper, we refer to the SSM obtained by
using FRANEC code \cita{Franec}, including: Livermore 2006 equation
of state (EOS) \cita{eos}; Livermore radiative opacity  tables
(OPAL) \cita{OPAL} calculated for the AS05 chemical composition; 
molecular opacities from ref. \cita{Ferguson05}  and 
conductive opacities from ref. \cita{Potekhin06} 
both calculated for AS05 composition;
nuclear reaction rates from NACRE compilation \cita{NACRE}, taking
into account the recent revision of the astrophysical factors
$S_{1,14}, S_{3,4}$ and $S_{1,7}$\cita{s114,s34,s17} and the $^7{\rm Be}$
electron capture rate from \cita{se7}.

\section{Linear expansion of the structure equations}
\label{Linear}

A variation of the input parameters (EOS, opacities, cross sections, etc.) 
with respect to the standard assumptions produces a solar model which deviates from SSM predictions.
The physical and chemical properties of the "perturbed" sun
can be described according to:
\begin{eqnarray}
\nonumber
h(r)&=&\overline{h}(r)[1+\delta h(r)]\\
\nonumber
X_i(r)&=&\overline{X}_i(r)[1+\delta X_i(r)]\\
Y(r)&=&\overline{Y}(r)+\Delta Y(r)
\end{eqnarray}
where $h = l,\,m,\,T,\,P,\,\rho$ and we use, here and in the following, the
notation $\overline{Q}$ to indicate the SSM prediction for the generic quantity $Q$.
For convenience, the deviations from SSM
are expressed in terms of relative variations $\delta h(r)$ and $\delta X_{i}(r)$ for all quantities
except the helium abundance, for which it is more natural
to use the absolute variation $\Delta Y(r)$.

In this paper, we will be mainly concerned with modifications of the radiative opacity, $\kappa(\rho,T,X_i)$, and 
of the energy generation coefficient, $\epsilon(\rho,T,X_i)$.
We indicate with $\delta \kappa(r)$ and $\delta \epsilon(r)$ the
relative variations of these quantities {\em along the SSM profile}, i.e.:
\begin{equation}
I(\overline{T}(r),\overline{\rho}(r),\overline{Y}(r),\overline{X}_{i}(r))=
\overline{I}(\overline{T}(r),\overline{\rho}(r),\overline{Y}(r),\overline{X}_{i}(r))\;[1+\delta I(r)]
\end{equation}
where $I=\kappa,\,\epsilon$.
When we consider the effect of a perturbation $\delta I(r)$ on the sun,
we have to take into account that the perturbed solar model, due to that modification, has
a different density, temperature and chemical composition with respect to the SSM.
The total difference $\delta I^{\rm tot}(r)$ between the perturbed sun and the
SSM {\em at a given point $r$} is defined by:
\begin{equation}
I(T(r),\rho(r),Y(r),X_{i}(r))=
\overline{I}(\overline{T}(r),\overline{\rho}(r),\overline{Y}(r),\overline{X}_{i}(r))[1+\delta I^{\rm tot}(r)].
\end{equation}
If we consider small perturbations ($\delta I \ll \overline{I}$), we can
expand to first order in $\delta T(r)$, $\delta \rho(r)$, $\delta X_i(r)$ and $\Delta Y(r)$, obtaining:
\begin{equation}
\delta I^{\rm tot}(r) = I_T(r)\, \delta T(r) + I_\rho(r)\,\delta\rho(r) +
I_Y(r)\, \Delta Y(r) +  \sum_i I_i(r)\, \delta X_i(r) + \delta I(r)
\label{deltaItot}
\end{equation}
Here,
the quantities $I_j$ describe the dependence of the properties of the
stellar plasma from temperature, density and chemical composition and are given by:
\begin{eqnarray}
\nonumber
I_{\rm \rho}(r)&=&\left.\frac{\partial \ln I}{\partial \ln \rho}\right|_{\rm SSM}\\
\nonumber
I_{\rm T}(r)&=&\left.\frac{\partial \ln I}{\partial \ln T}\right|_{\rm SSM}\\
\nonumber
I_{i}(r)&=&\left.\frac{\partial \ln I}{\partial \ln X_i}\right|_{\rm SSM}\\
I_{\rm Y}(r)&=&\left.\frac{\partial \ln I}{\partial Y}\right|_{\rm SSM}
\end{eqnarray}
where $I=\epsilon,\, \kappa,\, P$. The symbol $|_{\rm SSM}$
indicates that we calculate the derivatives $I_{j}$ along the density, temperature and chemical
composition profiles predicted by the SSM.

 By using the above notations,
we linearize the structure eqs. (\ref{eq5}), obtaining:
\begin{eqnarray}
\nonumber
\frac{\partial\, \delta m}{\partial r}  &=& \frac{1}{l_m} \,[\delta \rho - \delta m] \\
\nonumber
\frac{\partial\, \delta P}{\partial r}  &=& \frac{1}{l_P} \,[\delta m + \delta \rho - \delta P] \\
\nonumber
{\delta P}  &=& [P_\rho\,\delta \rho + P_T \,\delta T + P_Y \Delta Y +\sum_i P_i\, \delta X_i] \\
\nonumber
\frac{\partial\, \delta l}{\partial r}  &=& \frac{1}{l_l} \,[(1+\epsilon_\rho)\delta \rho +\epsilon_T \,\delta T +\epsilon_Y \Delta Y +
\sum_i\epsilon_i\, \delta X_i - \delta l + \delta \epsilon] \\
\nonumber
\frac{\partial\, \delta T}{\partial r}  &=& \frac{1}{l_T} \,[\delta l +(\kappa_T-4) \delta T + (\kappa_\rho+1) \delta\rho+ \kappa_Y \Delta Y
+\sum_i \kappa_i\, \delta X_i + \delta \kappa]\;\;\;\;\;\;\;\;{\rm Rad.}\\
\frac{\partial \, \delta T}{\partial r}  &=& \frac{1}{l_T} \,[\delta m + \delta \rho - \delta P] \;\;\;\;\;\;\;\;\;\;\;\;\;\;\;\;
 \;\;\;\;\;\;\;\;\;\;\;\;\;\;\;\; \;\;\;\;\;\;\;\;\;\;\;\;\;\;\;\; \;\;\;\;\;\;\;\;\;\;\;\;\;\;\;\;
 \;\;\;\;\;\;\;\;\;\;{\rm Conv.}
\label{linsyst}
\end{eqnarray}
where $l_h =\left[d\ln( \overline{h})/dr\right]^{-1}$ represents the scale height of the physical
parameter $h$ in the SSM.
The last two equations correspond to the linear expansion of the
energy transport equation in the radiative and in the convective regime respectively\footnote{In the presence of convection, 
we assume that $\nabla \equiv \nabla_{\rm ad}$
and that the adiabatic gradient is not affected by the performed modifications of the input parameters.}.
In our approach, we use the radiative transport equation for $r \le \overline{R}_{\rm b}$
and the convective transport equation for  $r > \overline{R}_{\rm b}$, 
where $\overline{R}_{\rm b}=0.730 \, R_{\odot}$ is the lower radius of the convective envelope predicted by our SSM.
We calculate {\it a posteriori}
the relative variation $\delta R_{\rm b}$ of the extension of the convective region
by applying the Schwarzchild criterion to the solutions of eqs.(\ref{linsyst}).
By expanding to first order, we obtain:
\begin{equation}
\delta R_{\rm b} = -\frac{\delta \nabla_{\rm rad, b}}{\zeta_{\rm b}} =  -\frac{\delta \kappa_{\rm tot, b}  + \delta P_{\rm b} + \delta l_{\rm b} - 4 \delta T_{\rm b} -\delta m_{\rm b}}
{\zeta_{\rm b}}
\label{ConvRadius}
\end{equation}
where $\zeta_{\rm b}  = d \ln \overline{\nabla}_{\rm rad}(\overline{R}_{\rm b})/d\ln r= 11.71$.
Here and in the following, the subscript ``b'' indicates that the various quantities are evaluated 
at $r=\overline{R}_{\rm b}$, i.e. at the bottom of the SSM's convective envelope.

\section{Properties of the sun and of the solar matter}
\label{Properties}

\begin{figure}[t]
\par
\begin{center}
\includegraphics[width=10.5cm,angle=0]{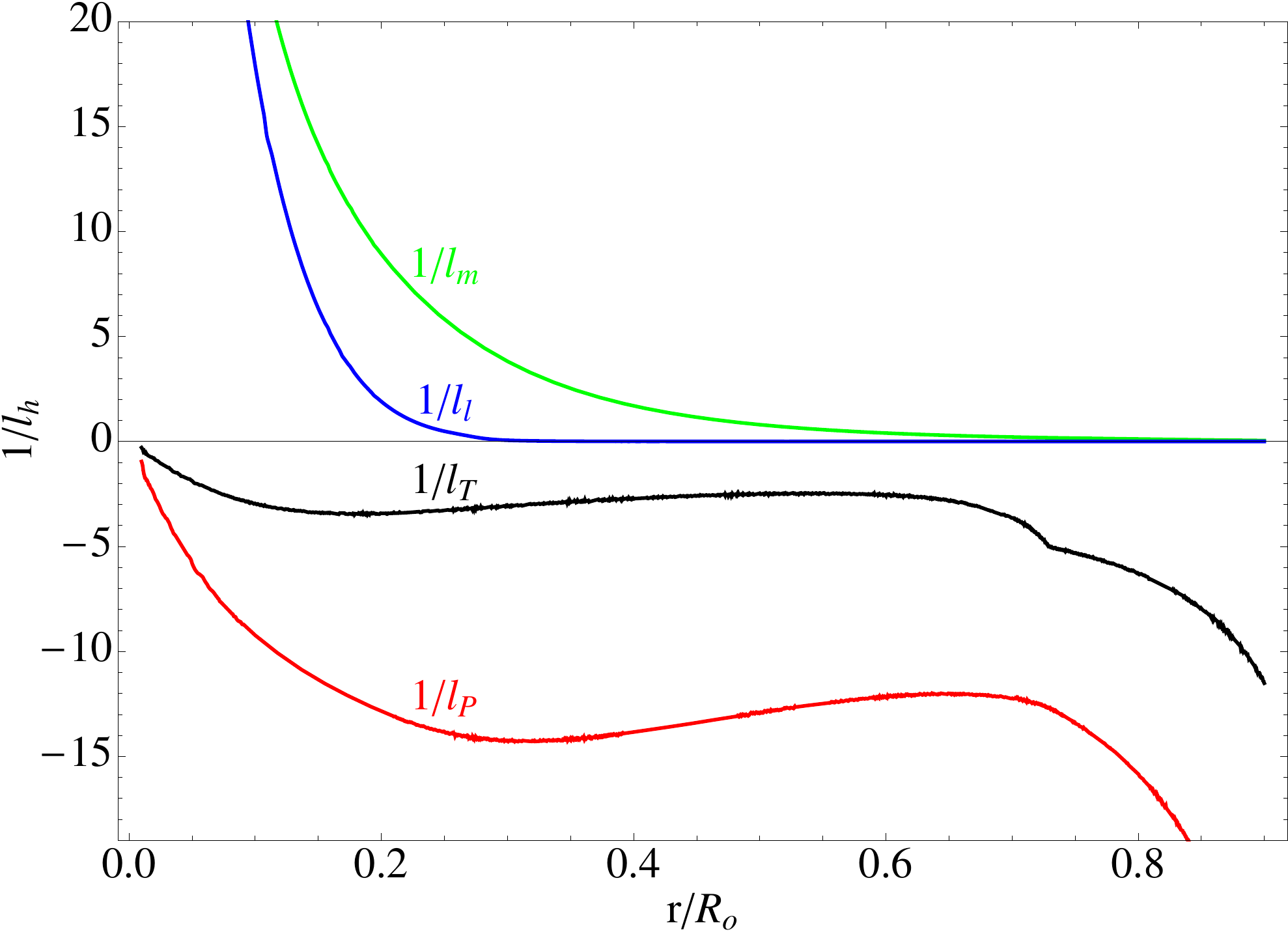}
\end{center}
\par
\vspace{-5mm} \caption{\em {\protect\small The inverse scale heights $1/l_{h}(r)$ of temperature (black), pressure (red), mass (green) and luminosity (blue) 
as a function of the solar radius.}}
\label{figlh}
\end{figure}

In order to define the linearized structure equations, we have to calculate
the functions $l_h(r)$ and the logarithmic derivatives $I_j(r)$.
In Fig.\ref{figlh}, we show the inverse scale heights $1/l_h(r)$ as a function of the solar radius.
The plotted results have been obtained numerically and refer to our SSM.
We see that the functions $1/l_P(r)$ and $1/l_T(r)$ vanish at the center of the sun
while they grow considerably (in modulus) in the outer regions, as a result of the
fast decrease of pressure and temperature close to surface.
The kink in the temperature scale height at $\overline{R}_{\rm b}= 0.730 R_{\odot}$ marks the transition from
the internal radiative region to the outer convective envelope.
The functions $1/l_m(r)$ and $1/l_l(r)$ have the opposite behaviour; they diverge at the solar center and
vanish at large radii, as it is expected by considering that most of the mass and of the
energy generation in the sun is concentrated close to the solar center.
To be more quantitative, the solar luminosity
is produced in the inner radiative core ($r\le 0.3\, R_{\odot}$), which contains approximately $60\%$
of the total solar mass. The radiative region ($r \le \overline{R}_{\rm b} $) which covers
about $40\%$ of the total volume of the sun, includes approximately $98\%$
of the total solar mass.


\begin{figure}[t]
\par
\begin{center}
\includegraphics[width=8.5cm,angle=0]{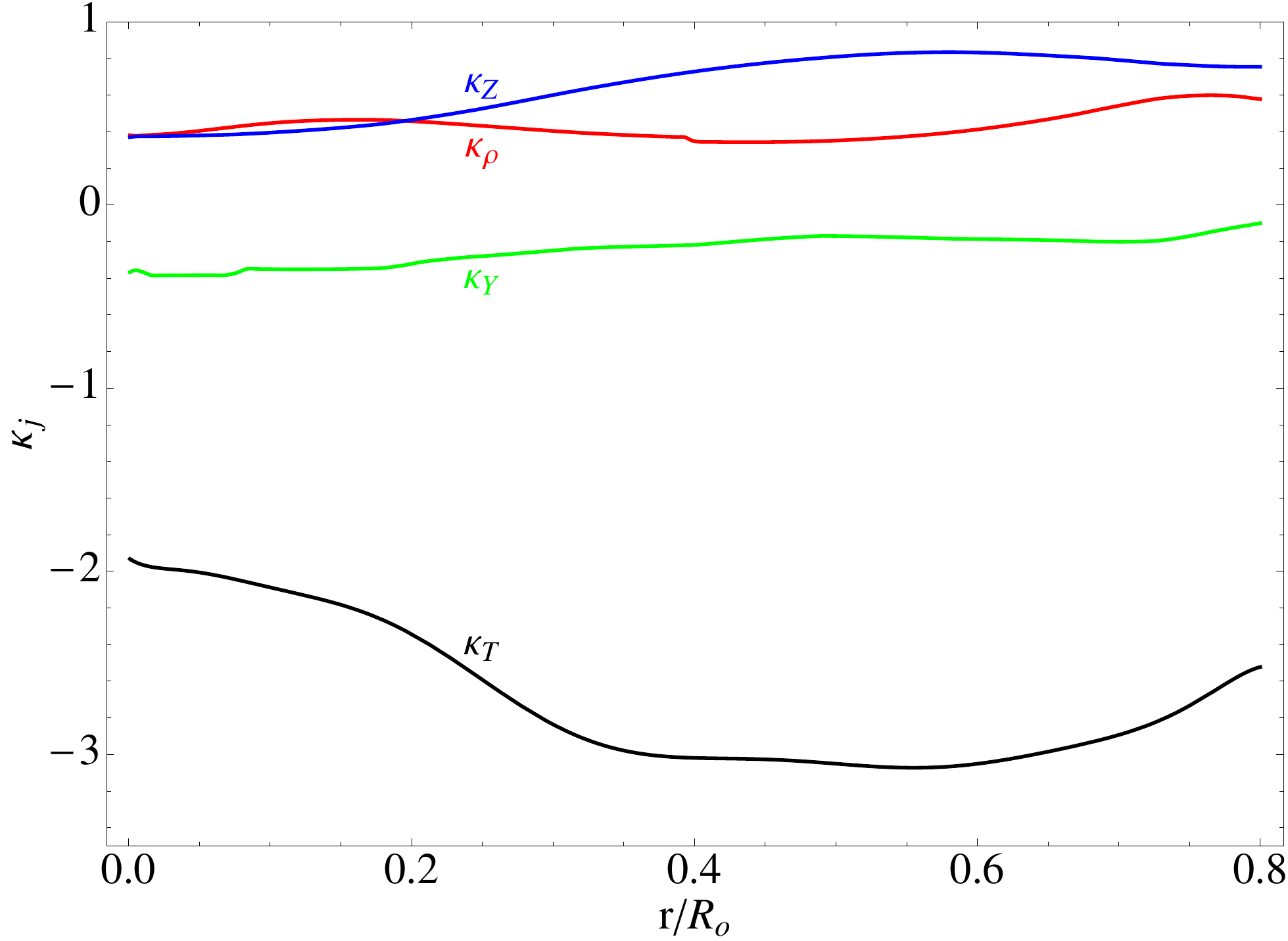}
\includegraphics[width=8.5cm,angle=0]{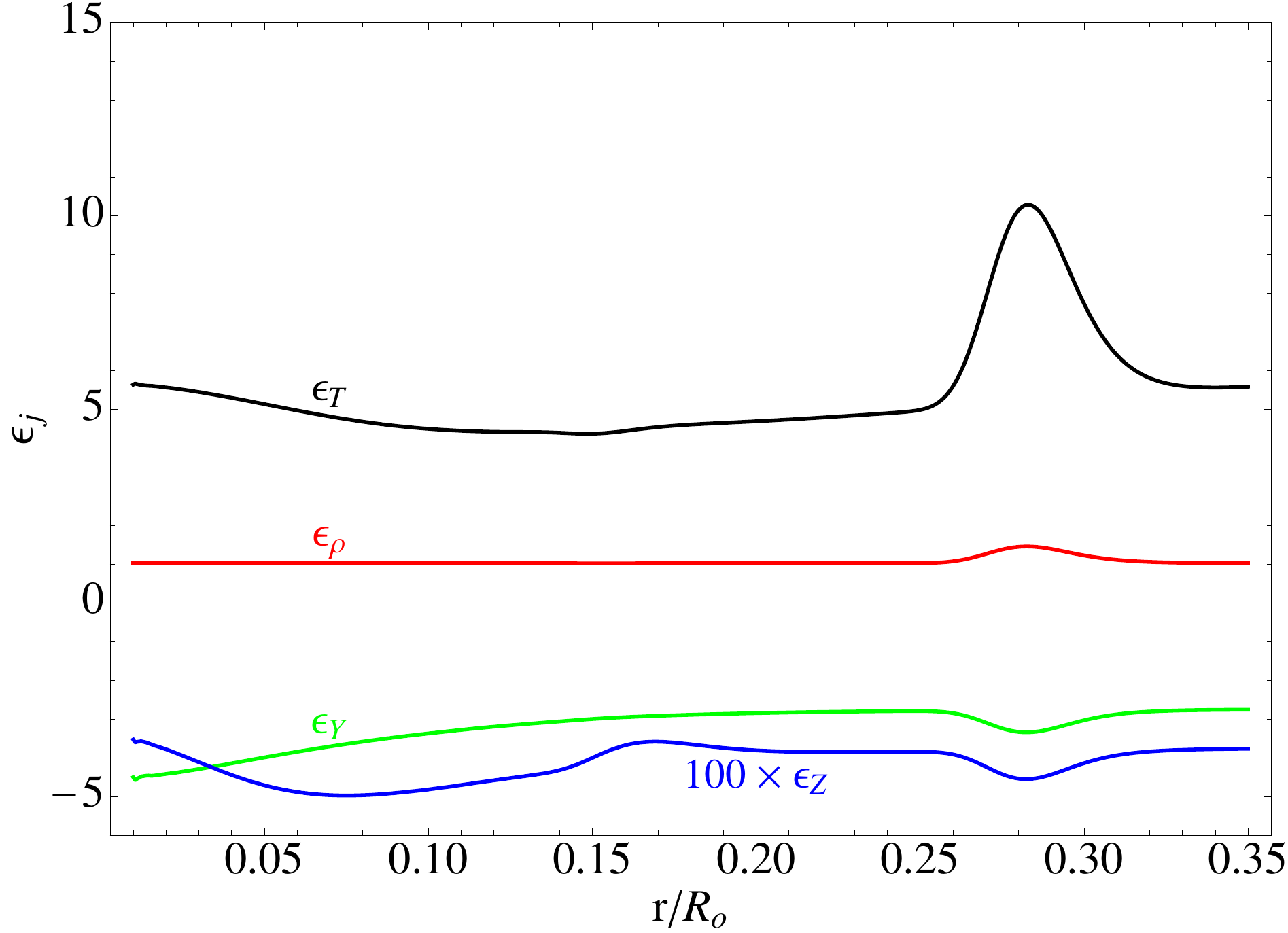}
\end{center}
\par
\vspace{-5mm} \caption{\em {\protect\small Left panel: The logarithmic derivative $\kappa_j(r)$ of the radiative opacity
with respect to temperature (black), density (red), helium (green) and metals (blue) as a function of the solar radius.
Right panel: The logarithmic derivative $\epsilon_j(r)$ of the energy generation coefficient
with respect to temperature (black), density (red), helium (green) and metals (blue - multiplied by a factor 100) as a function of the solar radius. }}
\label{figkappaepsilon}
\end{figure}

In Fig.\ref{figkappaepsilon}  we show the logarithmic derivatives $I_j(r)$
calculated numerically along the temperature, density and chemical composition profiles predicted by our SSM.
The left panel refers to the opacity derivatives $\kappa_j(r)$ and shows that opacity is a decreasing function of
temperature and helium abundance, while it is an increasing function of density.
The coefficient $\kappa_Z(r)\equiv \partial \ln \kappa/\partial \ln Z |_{\rm SSM}$
quantifies the dependence of opacity on the total metallicity $Z$. It
has been calculated by rescaling all the heavy element abundances
by a constant factor, so that
the metal admixture remains fixed. Metals provide about $\sim 40\%$
of the opacity at the center of the sun, while they account for
about $80\%$ of the total opacity at the bottom of the convective
region (see also \cita{Basu:2007fp}).

In the right panel of Fig.\ref{figkappaepsilon}, we present the logarithmic derivatives  $\epsilon_j(r)$
of the energy generation coefficient.
These have been calculated by assuming that the abundances of the
secondary elements for PP-chain and CN-cycle,
namely $^3 {\rm He}$, $^{12} {\rm C}$ and $^{14} {\rm N}$
 can be estimated as it is described in Appendix A.
The bumps at $r\simeq 0.28 \, R_{\odot}$ are related to
out-of-equilibrium behaviour of the $^3 {\rm He}$ abundance.
The presence of these features has, however, a negligible influence on the
solutions of eqs.($\ref{linsyst}$). The 
function $1/l_{l}(r)$, which multiplies all the $\epsilon_{j}(r)$ coefficients, 
drops, in fact,  rapidly to zero at $r \simeq 0.25 R_{\odot}$.\footnote{ 
For most applications, one could assume that secondary elements have the
equilibrium abundances all over the sun. The errors implied by this assumption are usually
negligible. In the region where it is not valid, in fact, the nuclear reaction rates are small 
and, thus, a negligible amount of energy and neutrinos are produced. 
In some cases (see sect.\ref{Neutrino}), however, it is necessary to improve this approximation.
 For this reason, we developed the method described 
in Appendix A, which allows to describe with good accuracy the behaviour of secondary elements
all over the sun. }

In the more internal region (where $^{3}{\rm He}$ assumes
the equilibrium value),  the displayed results can be understood by considering that energy is
produced by the PP-chain ($\sim 99\%$) with a small
contribution ($\sim 1\%$) by the CN-cycle.
The slight increase of $\epsilon_T(r)$ at $r \le 0.15\, R_{\odot}$
reflects the fact that the CN-cycle, which strongly depends on temperature,
gives a non-negligible contribution only at the center of the sun.
The coefficient $\epsilon_\rho(r)$ is approximately equal to one,
as it is expected by considering that the rate of two-body reactions in the unit mass
is proportional to the density.
Finally, the behavior of $\epsilon_Y(r)$ and $\epsilon_{Z}(r)$ is understood by considering
that the $pp$-reaction rate is proportional to the hydrogen abundance squared, i.e.
to $X^2=(1-Y-Z)^2$.


We have also evaluated numerically the logarithmic derivative of the pressure $P_{j}(r)$
by using the Livermore 2006 EOS \cita{eos}.
The obtained results can be approximated with good accuracy (about 1\%) by using
the perfect gas EOS: $ P=(\rho k_{\rm B}T)/(\mu \, m_{\rm u})$, where $m_{\rm u}$ is the atomic mass unit, $k_{\rm B}$ is the Boltzmann constant and
$\mu=(2 - 5/4\,Y - 3/2 \,Z)^{-1}$ is the mean molecular weight.
One obtains:
\begin{eqnarray}
\nonumber
P_{\rho}(r)&=&1\\
\nonumber
P_{\rm T}(r)&=&1\\
P_{Y}(r) &=& -\frac{\partial \ln \mu}{\partial Y}
=-\frac{5}{8-5Y(r)-6Z(r)}
\end{eqnarray}
The derivative with respect to the metal abundance
$P_{Z}(r)$ is of the order of few per cent and will be neglected in the following.
 By using the above relations, we obtain the following equation:
\begin{equation}
\delta \rho(r) = \delta P(r) -\delta T(r) - P_{Y} \Delta Y(r)
\end{equation}
that can be used to eliminate the quantity $\delta \rho$ from the linear structure eqs.(\ref{linsyst}).
We arrive at:
\begin{eqnarray}
\nonumber
\frac{d\delta m}{dr} &=& \frac{1}{l_m} \,\left[ \delta P - \, \delta T - \delta m
- P_Y \, \Delta Y  \right]\\
\nonumber
\frac{d\delta P}{dr}&=& \frac{1}{l_P}\, \left[ - \delta T + \delta m
- P_Y \, \Delta Y \right]\\
\nonumber
\frac{d\delta l}{dr}& = & \frac{1}{l_l}\, \left[ \beta_P \,\delta P + \beta_T \,\delta T - \delta l +
  \beta_Y\,\Delta Y +\beta_Z\,\delta Z +\delta \epsilon \right]\\
\nonumber
\frac{d\delta T}{dr}&=& \frac{1}{l_T}\, \left[ \alpha_P \,\delta P +
\alpha_T \,\delta T +
      \delta l + \alpha_Y\,\Delta Y  + \alpha_Z \delta Z +\delta \kappa \right]\;\;\;\;\;\;\;\;\;\;\;\;\;\;\;\;
 \;\;\;\;\;\;\;{\rm Rad.}\\
\frac{d\delta T}{dr}&=& \frac{1}{l_T}\, \left[ - \delta T + \delta m
- P_Y \, \Delta Y \right] \;\;\;\;\;\;\;\;\;\;\;\;\;\;\;\;
 \;\;\;\;\;\;\;\;\;\;\;\;\;\;\;\; \;\;\;\;\;\;\;\;\;\;\;\;\;\;\;\;
 \;\;\;\;\;\;\;\;\;\;{\rm Conv.}
 \label{linsyst2}
\end{eqnarray}
where the factors $\alpha_h$ and $\beta_h$ are given by:
\begin{eqnarray}
\nonumber
\alpha_P &=& \kappa_\rho + 1 \;\;\;\;\;\;\;\;\; \alpha_T = \kappa_T  - \kappa_\rho - 5 \;\;\;\;\;\;\;\;\;
\alpha_Y = - (\kappa_\rho +1)P_ Y + \kappa_Y \;\;\;\;\;\;\;\;\; \alpha_Z = \kappa_Z \\
\beta_P &=& \epsilon_\rho + 1 \;\;\;\;\;\;\;\;\;\; \beta_T = \epsilon_T - \epsilon_\rho - 1 \;\;\;\;\;\;\;\;\;\;
\beta_Y = -(\epsilon_\rho + 1)P_Y + \epsilon_Y \;\;\;\;\;\;\;\;\; \beta_Z = \epsilon_Z
\end{eqnarray}
In the above equations, we indicate with $\delta Z$ the relative variation
of the total metallicity, defined by:
\begin{equation}
Z(r)=\overline{Z}(r)\;[1+\delta Z(r)]
\end{equation}
and we implicitly assume that the heavy element admixture is
unchanged.\footnote{The effects of changes in the heavy element
admixture can be described in our approach by proper variations
$\delta k(r)$ and $\delta \epsilon(r)$ (see \cita{noiNext}).}


\section{Boundary conditions}
\label{Boundary}

In order to solve the linear structure equations, one has to specify the integration conditions.
This can be done quite easily at the center of the sun.
From the definitions of mass and luminosity, we know that $m(0)=0$ and $l(0)=0$.
This implies that:
\begin{equation}
1+\delta h(0)=\lim_{r\rightarrow0} \frac{h(r)}{\overline{h}(r)}= \frac{\frac{dh(0)}{dr}}{\frac{d\overline{h}(0)}{dr}}
\end{equation}
for $h=m,\, l$, so that one obtains:
\begin{eqnarray}
\nonumber
\delta m(0) & = & \delta P_{0} - \delta T_{0} -  P_{Y,0} \, \Delta Y_{0} \\
\nonumber
\delta P(0) & = & \delta P_{0} \\
\nonumber
\delta T(0) &=& \delta T_{0} \\
\delta l(0) &=& \beta_{P,0} \, \delta P_{0} + \beta_{T,0}  \,\delta T_{0}
+ \beta_{Y,0}  \, \Delta Y_{0} + \beta_{Z,0} \, \delta Z_0 +\delta \epsilon_{0}
\label{central}
\end{eqnarray}
where the subscript "0'' indicates that the various quantities are evaluated at the center of the sun.

To implement surface conditions, we exploit the fact that the sun has
a convective envelope where the solution of the linearized structure equations
can be explicitly calculated, as it is explained in the following.

First, we take advantage of the fact that there are no energy producing processes
and that a negligible fraction of the solar mass is contained
in the convective region.
This implies that $1/l_l(r)=0$ and $1/l_m(r)\simeq 0$, so that we can integrate the energy generation and the
continuity equation. We obtain $\delta l(r)\equiv 0$ and $\delta m(r)\simeq 0$ where
we clearly considered that only solutions that reproduce the observed solar luminosity and solar
mass are acceptable.

Then, we consider that the chemical abundances are constant due to the very efficient convective mixing, so that
$\Delta Y(r) \equiv \Delta Y_{\rm b} $ and  $P_Y(r) \Delta Y_{\rm b} \equiv - \delta \mu_{\rm b}$, where $\delta \mu_{\rm b}$
is the relative variation of the mean molecular weight at the bottom of the convective region.
By taking this into account, we can rewrite the transport equation in the form:
\begin{equation}
\frac{\partial \, \delta u}{\partial r} = - \frac{\delta u}{l_T}
\end{equation}
where $\delta u$ is the fractional variation of squared isothermal sound speed $u = P/\rho = k_{\rm B} T/\mu m_{\rm u} $. 
The general solution of this equation is $\delta u(r) = \delta u_{\rm b}  \,[\overline{T}_{\rm b}/\overline{T}(r)]$
which shows that, in order to avoid $\delta u(r)$ to
``explode'' at the surface of the sun,
 it necessarily holds
$\delta u(r)  = \delta T(r) - \delta \mu_{\rm b} \simeq 0$ in the
internal layers of the convective envelope.

By using this result into the hydrostatic equilibrium equation,  we
obtain $\delta P(r) = \delta \rho(r) \simeq \delta C$ where $\delta
C$ is an arbitrary constant. Finally, by considering that $\delta
\rho(r)$ is approximately constant in the internal layers of the
convective region (where most of the mass of the convective envelope
is contained), we use the continuity equation to improve the 
condition $\delta m(r)\simeq 0$, obtaining $\delta m_{\rm b} = -
\overline{m}_{\rm conv} \, \delta C$, where $\overline{m}_{\rm conv}=\overline{M}_{\rm
conv}/M_{\odot}=0.0192$ is the fraction of solar mass contained in
the convective region, as resulting from our SSM.

In conclusion, we have:
\begin{eqnarray}
\nonumber
\delta m(\overline R_{\rm b}) &=& - \overline{m}_{\rm conv} \, \delta C \\
\nonumber
\delta P(\overline R_{\rm b})&=& \delta C \\
\nonumber
\delta T(\overline R_{\rm b})&=&  \delta \mu_{\rm b} = - P_{Y,\rm b} \, \Delta Y_{\rm b} \\
\delta l(\overline R_{\rm b}) &=& 0
\label{surface}
\end{eqnarray}
 The factor $\delta C$ is a free parameter that
cannot be fixed from first principles, since we cannot model exactly convection in the
outermost super-adiabatic region. It has basically the same role as mixing length parameter
in SSM construction.

\section{The chemical composition of the sun}
\label{chemical}

The chemical composition of the perturbed sun should be calculated
 by integrating the perturbed structure and chemical-evolution
equations starting from an ad-hoc chemical homogeneous ZAMS model.
However, this would lead to several complications in our simplified approach.
We prefer to use a simple approximate procedure that allows to estimate with sufficient
accuracy the helium and metal abundances of the modified sun, without requiring to
follow explicitly its time-evolution.

\subsection{Notations}
In order to quantify the relevance of the different mechanisms determining the
present composition of the sun, we express the helium and metal abundance
according to:
\begin{eqnarray}
\nonumber
Y(r)&=&Y_{\rm ini}\,\left[ 1+D_{Y}(r) \right] +Y_{\rm nuc}(r)\\
Z(r)&=&Z_{\rm ini}\,\left[ 1+D_{Z}(r) \right]
\end{eqnarray}
Here,  $Y_{\rm ini}$ and $Z_{\rm ini}$ are the initial values
which, as explained previously, are free parameters in solar model construction
and are adjusted in order to reproduce the observed solar properties.
In our SSM, we have $\overline{Y}_{\rm ini}= 0.2611$ and $\overline{Z}_{\rm ini}=0.0140$.
The terms $D_{Y}(r)$ and $D_{Z}(r)$ describe the effects of elemental diffusion.
 Finally, $Y_{\rm nuc}(r)$ represents the total amount of helium produced in
the shell $r$ by nuclear processes and can be calculated by integrating the rates of
the helium-producing reactions during the sun history.

\begin{figure}[t]
\par
\begin{center}
\includegraphics[width=8.5cm,angle=0]{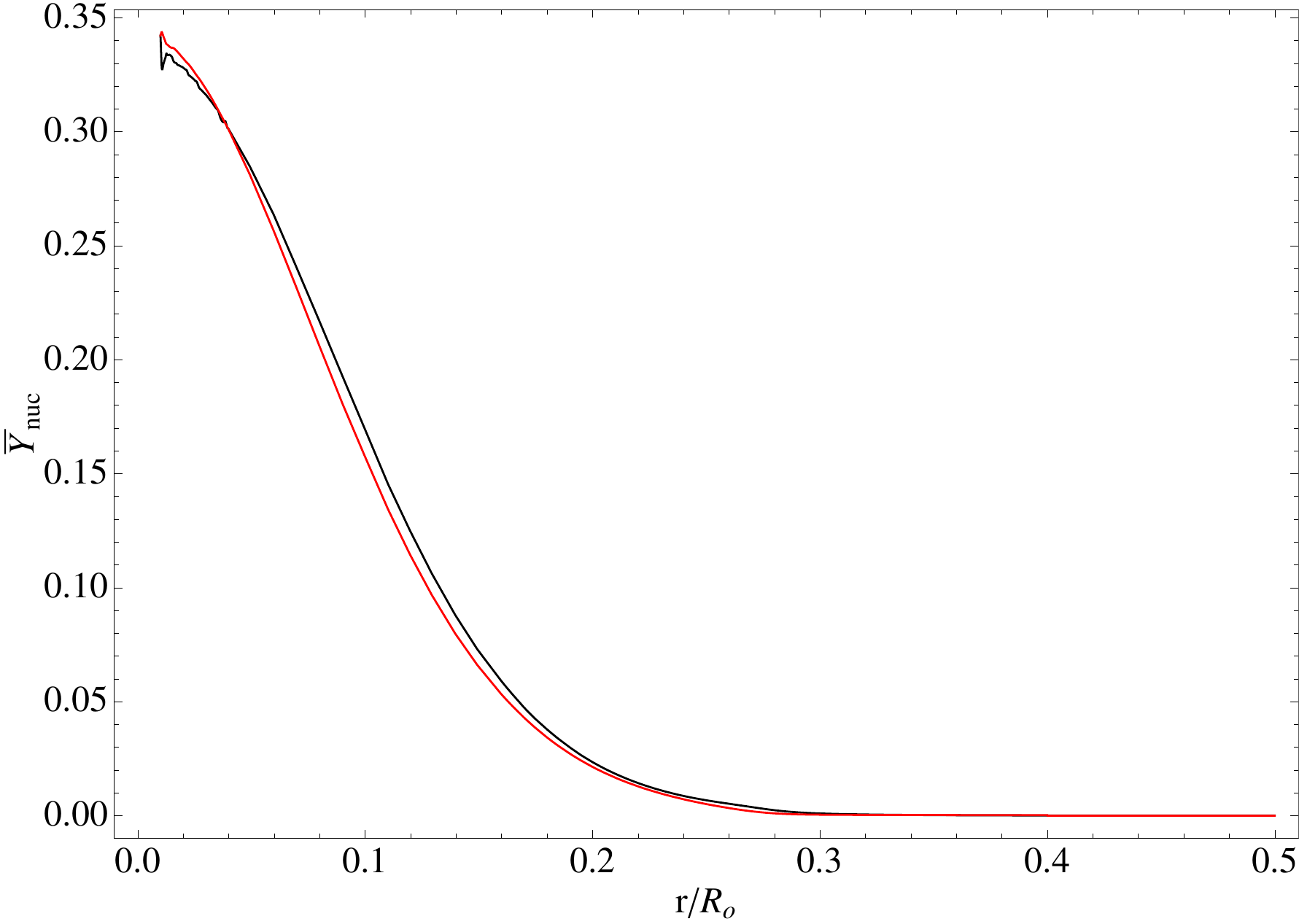}
\includegraphics[width=8.5cm,angle=0]{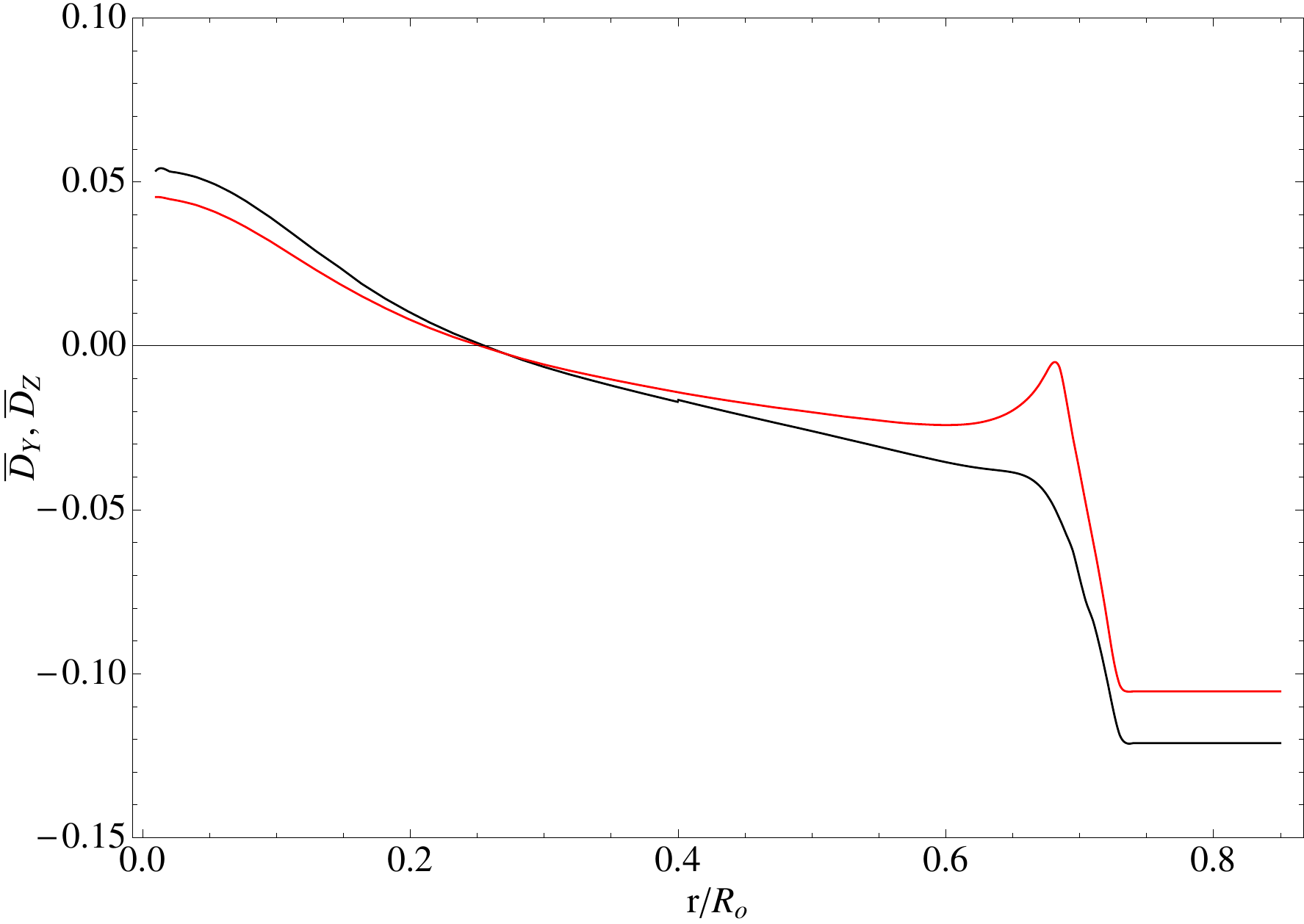}
\end{center}
\par
\vspace{-5mm} \caption{\em {\protect\small Left panel:  The black line shows the amount of helium produced by nuclear reactions $\overline{Y}_{\rm nuc}(r)$.  
The red line shows the energy generation coefficient $\overline{\epsilon}(r)$ suitably rescaled to superimpose with $\overline{Y}_{\rm nuc}(r)$.
Right panel: The terms $\overline{D}_{Y}(r)$ and $\overline{D}_{Z}(r)$ which describe the effects of elemental diffusion in the SSM.}}
\label{figynuc}
\end{figure}

In Fig.\ref{figynuc} we show the quantities $\overline{Y}_{\rm nuc}(r)$, $\overline{D}_{Y}(r)$ and
$\overline{D}_{\rm Z}(r)$ estimated from our SSM.
Nuclear helium production is relevant in the internal radiative
core ($r\le 0.3 R_{\odot}$) where it is responsible for an enhancement
of the helium abundance which can be as large as $\overline{Y}_{\rm nuc,0}\simeq 0.35$
at the center of the sun. Elemental diffusion
accounts for a $\sim 10\%$ increase of helium and metals in the
central regions with respect to external layers of the radiative region.
The convective envelope, due to the very efficient convective mixing, is
chemically homogeneous and can be fully described in terms of two numbers:
the surface helium and metal abundances, indicated with $Y_{\rm b}$  and $Z_{\rm b}$ respectively\footnote{
We use the subscript ``b'' to emphasize that the surface abundances coincide with the
 values at the bottom of the convective region.}. These are related to initial values by
\begin{eqnarray}
Y_{\rm b}&=&Y_{\rm ini}\,\left[ 1+D_{Y,\rm b} \right]\\
\nonumber
Z_{\rm b}&=&Z_{\rm ini}\,\left[ 1+D_{Z,\rm b} \right]
\end{eqnarray}
where $D_{Y,\rm b}$ and $D_{Z,\rm b}$ parametrize the effects of elemental diffusion.
In our SSM, we have $\overline{Y}_{\rm b}=0.229$, $\overline{Z}_{\rm b} = 0.0125$, 
$\overline{D}_{Y,\rm b}=-0.121$, $\overline{D}_{Z,\rm b}=-0.105$.


We are interested in describing how the chemical composition is modified when we perturb
the SSM.  In the radiative core ($r\le \overline{R}_{\rm b}$), 
we neglect the possible variations of the diffusion terms\footnote{The diffusion terms $D_{Y}(r)$ and $D_{Z}(r)$ are at the few per cent level
in the radiative region. Their variations are, thus, expected to produce very small effects on the solar composition.} 
and we write:
\begin{eqnarray}
\nonumber
\Delta Y(r) &=& \Delta Y_{\rm ini} \,\left[ 1+\overline{D}_{Y}(r) \right] +\Delta Y_{\rm nuc}(r) \\
\delta Z(r) &=& \delta Z_{\rm ini}
\label{ChimRad}
\end{eqnarray}
where $\Delta Y_{\rm nuc}(r)$ is the absolute variation of the amount of helium produced by
nuclear reactions. A better accuracy is required in the convective region, because the surface helium abundance $Y_{\rm b}$ 
is an observable quantity. We, thus, discuss explicitly the role of diffusion and we write:
\begin{eqnarray}
\nonumber
\Delta Y_{\rm b}&=& (1+\overline{D}_{Y, \rm b})\, \Delta Y_{\rm ini}  + \overline{Y}_{\rm ini}\, \overline{D}_{Y,\rm b} \, \delta D_{Y,{\rm b}}\\
\delta Z_{\rm b}&=&\delta Z_{\rm ini} +\frac{\overline{D}_{Z, \rm b}}{1+\overline{D}_{Z, \rm b}}\, \delta D_{Z,_{\rm b}}
\label{ChimSup}
\end{eqnarray}
where $\delta D_{Y,\rm b}$  and $\delta D_{Z, \rm b}$ are the fractional variations of the diffusion terms.

It is important to remark that $\Delta Y_{\rm b}$ and $\delta Z_{\rm b}$
are related among each other, since the metals-to-hydrogen ratio at the
surface of the sun is observationally fixed.  By imposing $\delta (Z/X)_{\rm b} = 0$, we obtain:
\begin{equation}
\delta Z_{\rm b}=-\frac{1}{1-\overline{Y}_{\rm b}}\Delta Y_{\rm b}
\label{Zsup}
\end{equation}
where we considered that $X_{\rm b} \simeq 1-Y_{\rm b}$. This relation can be rewritten in terms of the initial helium and metal abundances,
obtaining:
 \begin{equation}
\delta Z_{\rm ini} = Q_0 \, \Delta Y_{\rm ini}
+Q_1 \, \delta D_{Y,{\rm c}}
 +Q_2 \, \delta D_{Z,_{\rm c}}
\label{Zini}
\end{equation}
where the coefficients $Q_{i}$ are given by:
\begin{eqnarray}
\nonumber
Q_0  &=&  -\frac{1+\overline{D}_{Y, \rm b}}{1-\overline{Y}_{\rm b}} = - 1.141 \\
\nonumber
Q_1  &=& -\frac{\overline{Y}_{\rm ini}\, \overline{D}_{Y,\rm b}}{1-\overline{Y}_{\rm b}} = +0.041 \\
 Q_2 &=& -\frac{\overline{D}_{Z, \rm b}}{1+\overline{D}_{Z, \rm b}} = +0.118 \, .
\end{eqnarray}

\subsection{Production of helium by nuclear reactions}

\begin{figure}[t]
\par
\begin{center}
\includegraphics[width=8.5cm,angle=0]{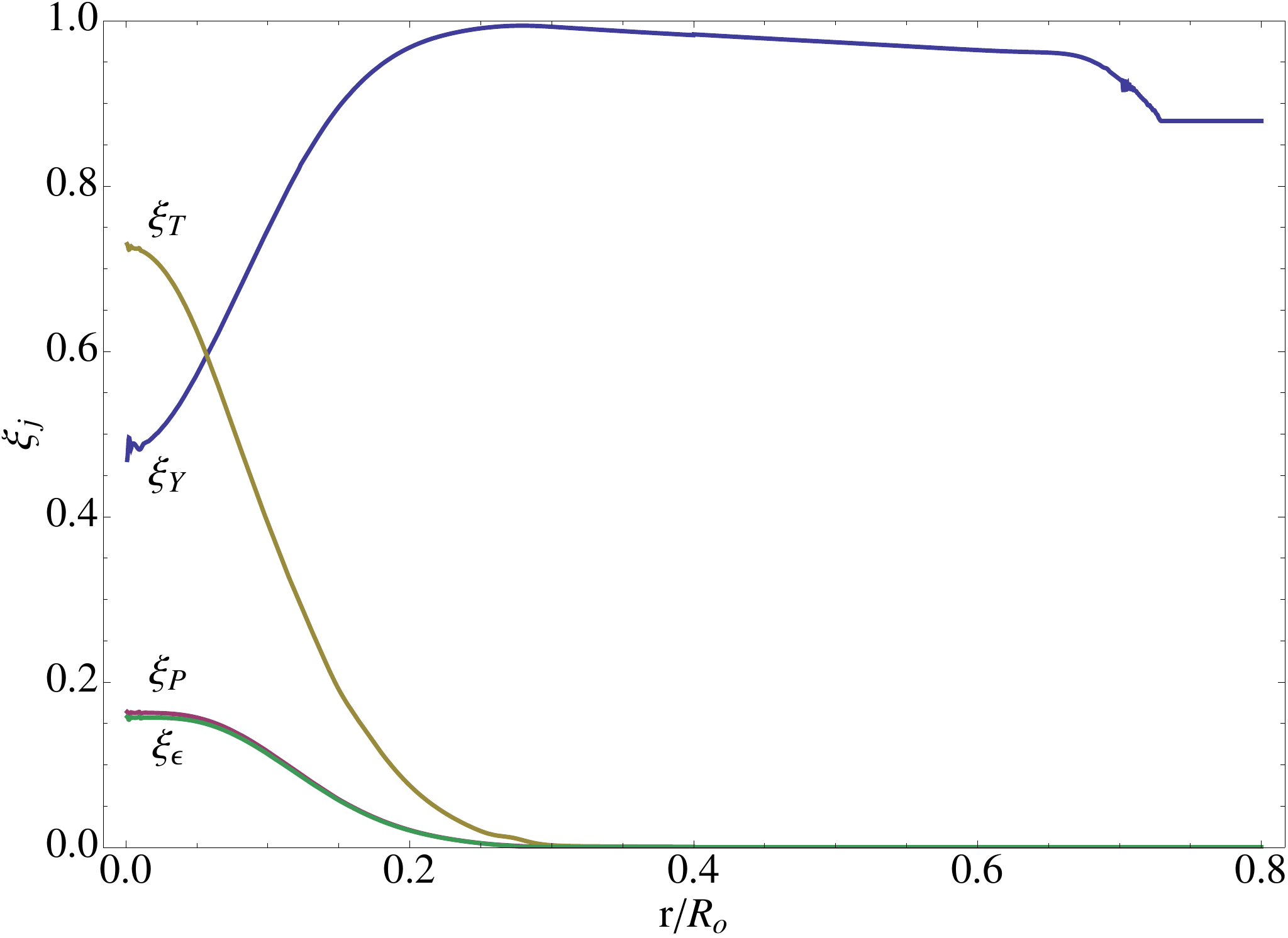}
\end{center}
\par
\vspace{-5mm} \caption{\em {\protect\small The factors $\xi_j(r)$ defined in eqs.(\ref{xij}).}}
\label{figxij}
\end{figure}

 In order to predict the helium abundance in the radiative region,
one has to estimate the variation of nuclear production of helium $\Delta Y_{\rm nuc}(r)$,
 see eq.(\ref{ChimRad}).
We note that the quantity ${\overline Y}_{\rm nuc}(r)$ varies proportionally to
$\overline{\epsilon}(r)$ in the SSM, as it is seen from Fig.\ref{figynuc}. This is natural since the
helium production rate is directly proportional to the energy generation rate and, moreover,
the energy generation profile
$\overline{\epsilon}(r)$ is nearly constant during the past history of the sun.
We assume that the observed proportionality holds true also in modified solar models,
obtaining as a consequence the following relation:
\begin{equation}
\Delta Y_{\rm nuc}(r) = \overline{Y}_{\rm nuc}(r) \delta \epsilon^{\rm tot}(r)
\end{equation}
If we expand $\delta \epsilon^{\rm tot}(r)$ according to rel.(\ref{deltaItot}) and solve with respect to
$\Delta Y(r)$, we can recast eq.(\ref{ChimRad}) in the form:\footnote{In the expansion of $\delta \epsilon_{\rm tot}(r)$ 
we neglected the term $\epsilon_Z(r)\delta Z(r)$. This term is expected to give a
negligible contribution since the coefficient $\epsilon_Z(r)$ is at most equal to -0.05, whereas the 
coefficients $\epsilon_{T}(r)$, $\epsilon_\rho(r)$ and $\epsilon_{Y}(r)$ are order unity or larger, see right panel of fig.\ref{figkappaepsilon}.}
\begin{equation}
\Delta Y(r)=\xi_Y(r)\,\Delta Y_{\rm ini}+\xi_P(r)\,\delta P(r)+\xi_T(r)\,\delta T(r)+\xi_{\epsilon}(r)\,\delta\epsilon(r)
\label{helium}
\end{equation}
where:
\begin{eqnarray}
\label{xij}
\nonumber
\xi_Y(r)  &=&\frac{1+\overline{D}_{Y}(r)}{1-\overline{Y}_{\rm nuc}(r) \left[\epsilon_Y(r)  -P_{Y}(r)\, \epsilon_{\rho}(r) \right]}\\
\nonumber
\xi_P(r) &=&\frac{\overline{Y}_{\rm nuc}(r)\,\epsilon_\rho(r)}{1-\overline{Y}_{\rm nuc}(r) \left[\epsilon_Y(r)  -P_{Y}(r)\, \epsilon_{\rho}(r) \right]}\\
\nonumber
\xi_T(r) &=&\frac{\overline{Y}_{\rm nuc}(r)\left[ \epsilon_T(r) - \epsilon_\rho(r)\right]}{1-\overline{Y}_{\rm nuc}(r) \left[\epsilon_Y(r)  -P_{Y}(r)\, \epsilon_{\rho}(r) \right]}\\
\xi_\epsilon(r) &=&\frac{\overline{Y}_{\rm nuc}(r)}{1-\overline{Y}_{\rm nuc}(r) \left[\epsilon_Y(r)  -P_{Y}(r)\, \epsilon_{\rho}(r) \right]}\;.
\end{eqnarray}
The factors $\xi_j(r)$ are shown in Fig.\ref{figxij}.


\subsection{Elemental diffusion in the convective region}

The total mass $M_{{\rm conv}, i}$ of the $i-$th element contained
in convective envelope evolves due to elemental diffusion according to relation \cita{aller,bahdif1}:
\begin{equation}
\frac{1}{M_{{\rm conv}, i}}\frac{d M_{{\rm conv}, i}}{dt}=\frac{\omega_{i}}{H}
\label{diffconv}
\end{equation}
Here, the parameter $H$ is the ``effective thickness'' of the convective region
defined by the relation $4\pi R_{\rm b}^2\, H \,\rho_{\rm b}=M_{\rm conv}$.
The quantity  $\omega_{i}$ is the diffusion velocity of the $i-$th element
 at the bottom of the convective region
that can be expressed as:
\begin{equation}
\label{omega}
\omega_{i}=\frac{T_{\rm b}^{5/2}}{\rho_{\rm b}} \left[(A_{P,i}+\nabla_{\rm ad}  \, A_{T,i})\frac{\partial \ln P(R_{\rm b})}{\partial r} \right]
\end{equation}
where  $A_{P, i }$, $A_{T, i }$ are the diffusion coefficients (see \cita{bahdif1} for details).
In the above relation,
we neglected, for simplicity, the term proportional to the hydrogen concentration
gradient\footnote{This term gives a sub-dominant contribution at the present stage and, thus,
is negligible also in the initial phases of the evolution when the sun is essentially homogeneous}
and we took advantage of the fact that $\partial \ln T /\partial r  = \nabla_{\rm ad}\,\partial \ln P /\partial r$
at the bottom the convective region, as it is prescribed by the Schwartzchild criterion.

We have to estimate the effect of a generic modification of the SSM on elemental diffusion.
In most cases, a good description is obtained by assuming that $D_{i,\rm b}$
varies proportionally to the efficiency of diffusion in the present sun, i.e. to the the r.h.s of eq.(\ref{diffconv})
evaluated at the present time.
 This implies that:
\begin{equation}
\delta D_{i,\rm b} = \delta \omega_{i} - \delta H
\end{equation}
where $\delta H$ is the relative variation of the convective envelope effective thickness and
$\delta \omega_{i}$ is the relative variation of the diffusion velocity.
By following the calculations described in the appendix B,
 one is able to show that:
\begin{eqnarray}
\label{EQ_diffusion}
\nonumber
\delta D_{Y,\rm b} &=& \Gamma_{Y} \; \delta T_{\rm b} +  \Gamma_{P} \, \delta P_{\rm b}  \\
\delta D_{Z,\rm b} &=& \Gamma_{Z} \; \delta T_{\rm b} +  \Gamma_{P} \, \delta P_{\rm b}
\end{eqnarray}
where $\Gamma_{Y} = 2.05 $, $\Gamma_{Z}= 2.73$ and $\Gamma_{P}= -1.10$ and we assumed that
all metals have the same diffusion velocity as iron.

By using the above relations and taking into account eqs.(\ref{ChimSup},\ref{Zsup})
one obtains the surface abundances variations $\Delta Y_{\rm b}$ and $\delta Z_{\rm b}$ as a function of the free parameters $\Delta Y_{\rm ini}$
and $\delta C$. Namely, we obtain:
\begin{eqnarray}
\nonumber
\Delta Y_{\rm b} &=& A_{\rm Y} \; \Delta Y_{\rm ini} + A_{\rm C} \; \delta C \\
\label{ChimSurfFinal}
\Delta Z_{\rm b} &=& B_{\rm Y} \; \Delta Y_{\rm ini} + B_{\rm C} \; \delta C
\end{eqnarray}
where
\begin{eqnarray}
\nonumber
A_{\rm Y} & = &  \frac{1+\overline{D}_{Y, \rm b}} { 1 + P_{Y,\rm b} \, \overline{Y}_{\rm ini} \, \overline{D}_{Y, \rm b} \, \Gamma_{Y}} =0.838\\
A_{\rm C} & = &  \frac{\overline{Y}_{\rm ini} \, \overline{D}_{Y, \rm b} \, \Gamma_{P}} { 1 + P_{Y,\rm b} \, \overline{Y}_{\rm ini} \, \overline{D}_{Y, \rm b} \,\Gamma_{Y}} =0.033
\end{eqnarray}
and
\begin{eqnarray}
\nonumber
B_{\rm Y} & = &  - \frac{A_{\rm Y}} { 1 - \overline{Y}_{\rm b} } = -1.088 \\
B_{\rm C} & = &  - \frac{A_{\rm C}} { 1 -\overline{Y}_{\rm b} } = -0.043
\end{eqnarray}
Moreover, we can use rels.(\ref{Zini},\ref{surface}) to calculate the variation of the initial metal abundance $\delta Z_{\rm ini}$
which, in our approach, coincides with the variation of metal abundance in the radiative region $\delta Z(r)$. 
We obtain:
\begin{equation}
\delta Z(r) = \delta Z_{\rm ini} = Q_{\rm Y} \, \Delta Y_{\rm ini} + \, Q_{\rm C} \, \delta C
\label{MetalRad}
\end{equation}
where:
\begin{eqnarray}
\nonumber
Q_{\rm Y} &=& Q_0 -P_{Y,\rm b} \, A_{\rm Y}\,  \left(Q_1 \, \Gamma_{Y}  + Q_2 \, \Gamma_{Z} \right) = - 0.887\\
Q_{\rm C} &=& -P_{Y,\rm b}\, A_{\rm C} \,  \left( Q_1 \,\Gamma_{Y} + Q_2\,\Gamma_{Z}\right) + \Gamma_{P} \, \left( Q_1 + Q_2 \right) = -0.164
\end{eqnarray}
If we had neglected the effect of diffusion, we would have obtained $A_{\rm Y} = 0.879$ and $A_{\rm C} = 0$, $B_{\rm Y} =-1.141$ and $B_{\rm C}= 0 $
and $Q_{\rm Y} = Q_0 = - 1.14$ and $Q_{\rm C}=0$.


\begin{figure}[t]
\par
\begin{center}
\includegraphics[width=8.0cm,angle=0]{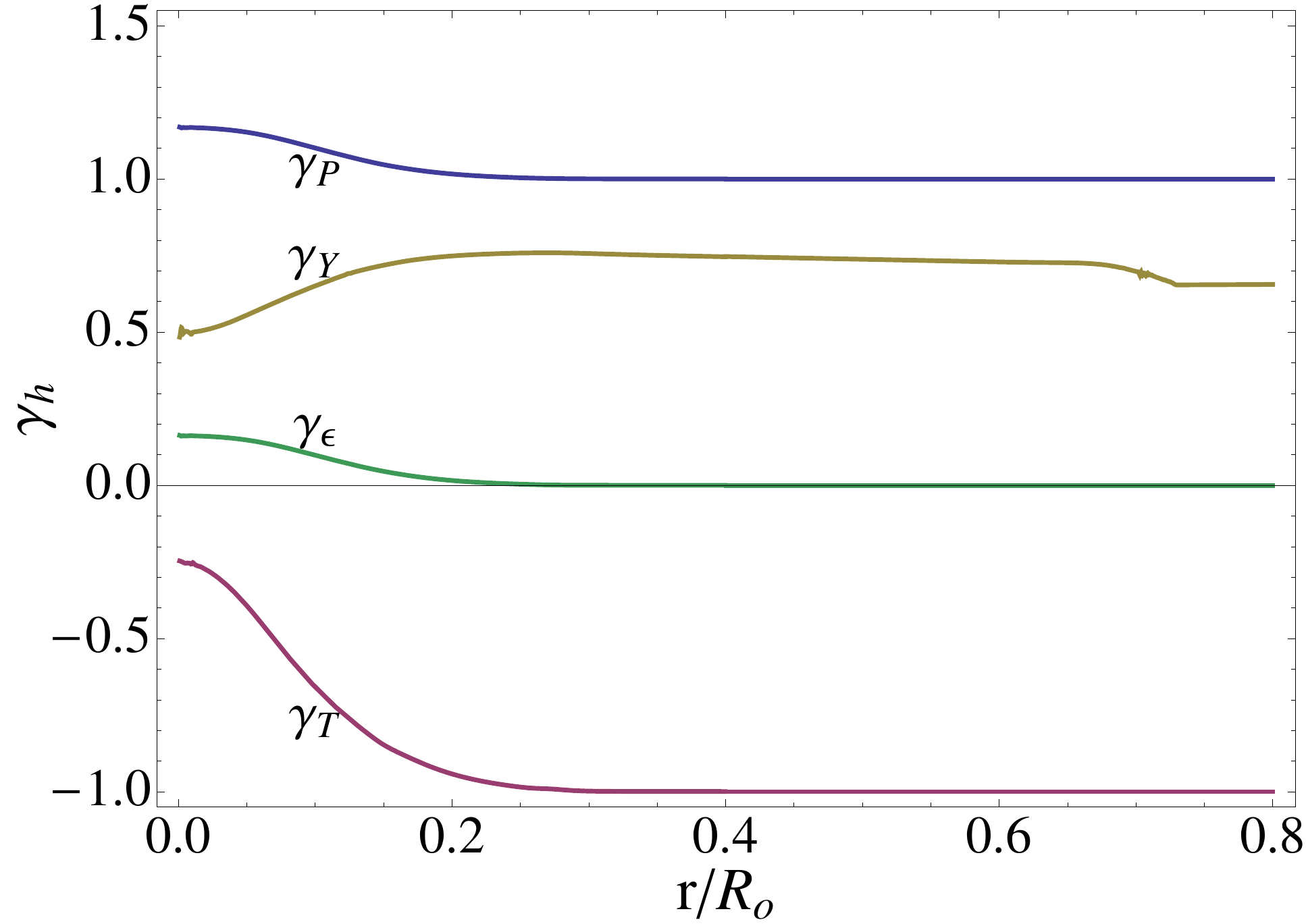}
\end{center}
\par
\vspace{-5mm} \caption{\em {\protect\small The coefficients $\gamma_{h}(r)$ which define the continuity and the hydrostatic equilibrium equation.}}
\label{HydroFinal}
\end{figure}


\section{The final set of equations}
\label{Fullset}

 We have now all the ingredients to formally define LSMs.
The equations obtained in Sect.\ref{chemical}, namely eqs.(\ref{helium}), (\ref{ChimSurfFinal}) and (\ref{MetalRad}),
relate the chemical composition of the modified sun to the {\em present} values of the structural parameters
$\delta P(r)$ and $\delta T(r)$, to the energy generation coefficient  $\delta \epsilon(r)$ and to the free parameters $\Delta Y_{\rm ini}$ and $\delta C$.
These relations can be inserted in eqs.(\ref{linsyst2})  and in the integration conditions, eqs.(\ref{central}) and (\ref{surface}),
obtaining a linear system of ordinary differential equations
that completely determines the physical and chemical properties of the modified sun.
 In this section, we give the equations in their final form.

\begin{figure}[t]
\par
\begin{center}
\includegraphics[width=8.0cm,angle=0]{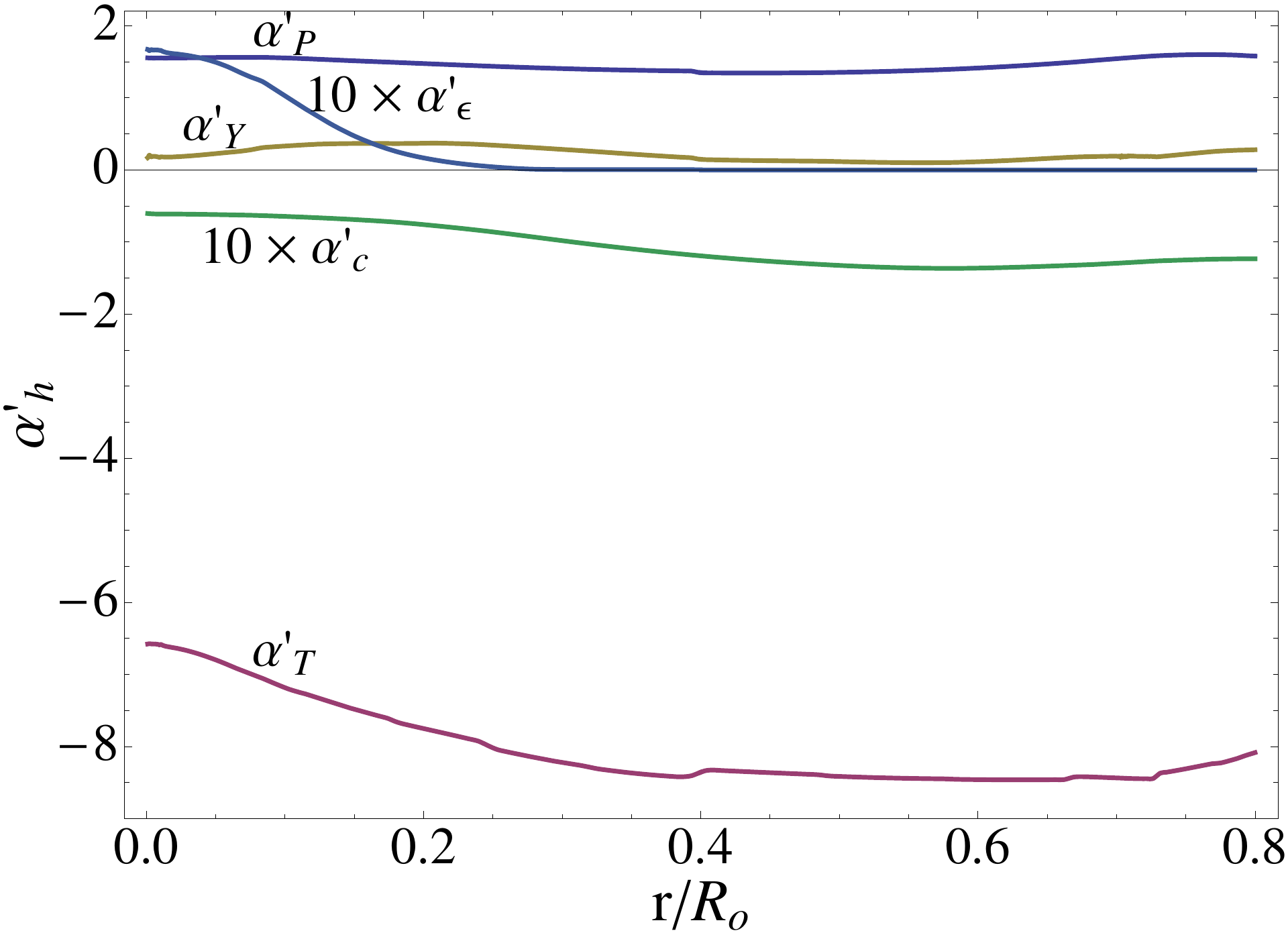}
\includegraphics[width=8.0cm,angle=0]{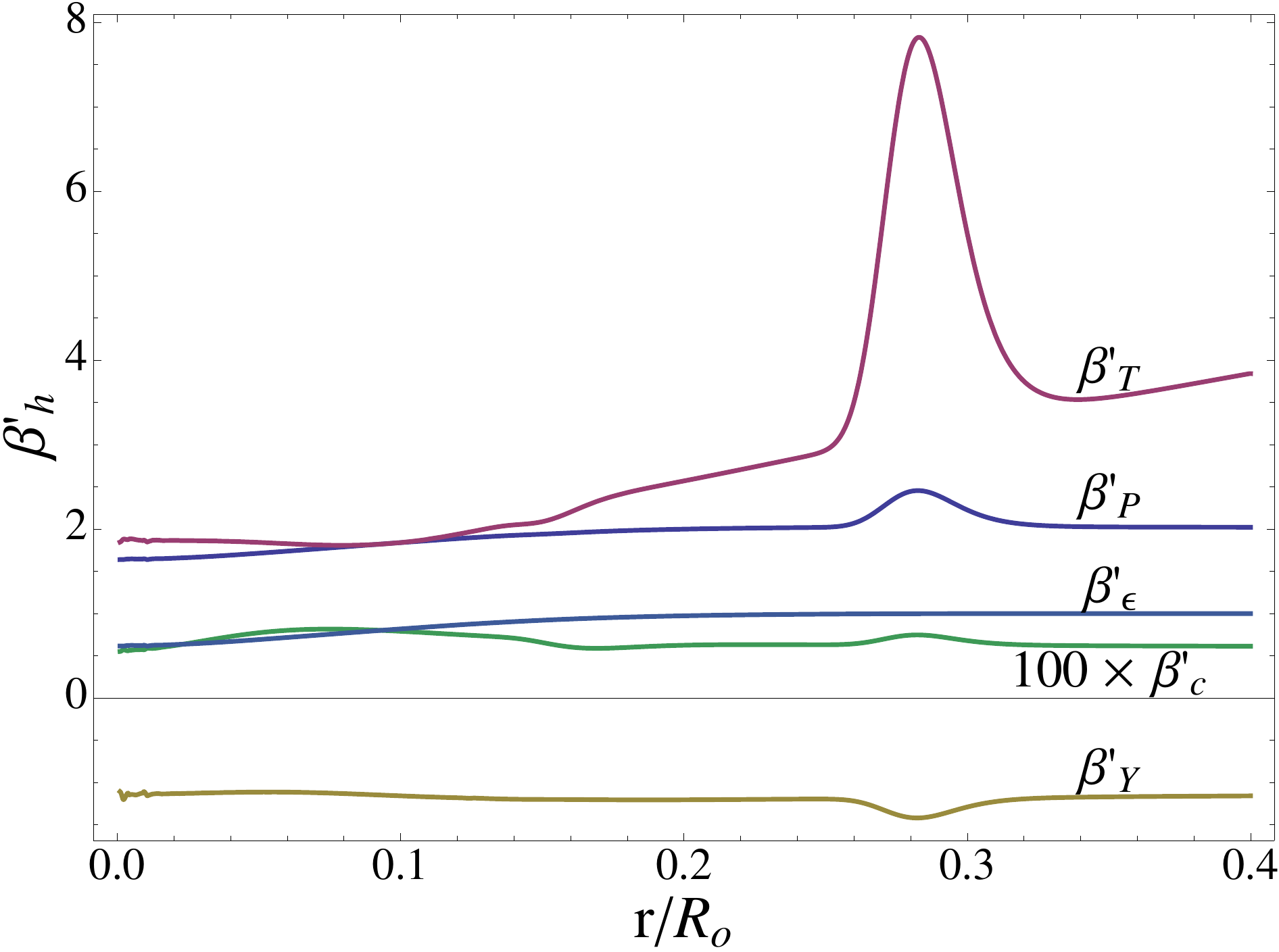}
\end{center}
\par
\vspace{-5mm} \caption{\em {\protect\small The coefficients $\alpha'_{h}(r)$ and $\beta'_{h}(r)$ which defines the energy transport and the energy conservation
equation.}}
\label{TransportEnergyFinal}
\end{figure}

The properties of the sun in the radiative region ($r\le \overline{R}_{\rm b}$) are described by:
\begin{eqnarray}
\nonumber
\frac{d\delta m}{dr} &=& \frac{1}{l_m} \,\left[ \gamma_{P} \, \delta P + \gamma_T \, \delta T - \delta m
+ \gamma_Y \, \Delta Y_{\rm ini}  + \gamma_\epsilon \; \delta \epsilon \right]\\
\nonumber
\frac{d\delta P}{dr}&=& \frac{1}{l_P}\, \left[(\gamma_{P} - 1) \, \delta P + \gamma_T \, \delta T + \delta m
+ \gamma_Y \, \Delta Y_{\rm ini}  + \gamma_{\epsilon} \; \delta \epsilon \right]\\
\nonumber
\frac{d\delta l}{dr}& = & \frac{1}{l_l}\, \left[ \beta'_P \,\delta P +  \beta'_T \,\delta T - \delta l +
 \beta'_Y \,\Delta Y_{\rm ini} + \beta'_C \, \delta C + \beta'_\epsilon \,\delta \epsilon \right]\\
\frac{d\delta T}{dr}&=& \frac{1}{l_T}\, \left[ \alpha'_P \,\delta P + \alpha'_T \,\delta T + \delta l +
\alpha'_Y \, \Delta Y_{\rm ini}+ \alpha'_C \, \delta C + \delta \kappa  + \alpha'_\epsilon \, \delta \epsilon \right]
\label{linsyst3}
\end{eqnarray}
The coefficients $\gamma_{h}$ which define the continuity and the hydrostatic equilibrium equation are given by:
\begin{equation}
\nonumber
\gamma_P = 1- P_Y \xi_P \;\;\;\;\;\;
\gamma_T = - 1 + P_Y \xi_T \;\;\;\;\;\;
\gamma_Y = - P_Y \xi_Y\, \;\;\;\;\;\;
\gamma_\epsilon =  -P_Y\xi_\epsilon
\end{equation}
and are shown in Fig.\ref{HydroFinal} as a function of the solar radius.
The coefficients $\alpha'_{h}$ and $\beta'_{h}$ which define the transport
and the energy generation equation are given by:
\begin{eqnarray}
\nonumber
&\beta'_P& = \beta_P + \beta_Y\xi_P \;\;\;\;\;\;
\beta'_T =  \beta_T + \beta_Y\xi_T \;\;\;\;\;\;
\beta'_Y =  \beta_Y\xi_Y +\beta_Z  Q_{\rm Y} \;\;\;\;\;\;
\beta'_C = \beta_Z Q_{\rm C}  \;\;\;\;\;\;
\beta'_\epsilon=  1+ \beta_Y \xi_{\epsilon} \\
&\alpha'_P& = \alpha_P + \alpha_Y\xi_P \;\;\;\;\;
\alpha'_T =  \alpha_T + \alpha_Y\xi_T \;\;\;\;\;
\alpha'_Y =  \alpha_Y\xi_Y +\alpha_Z  Q_{Y} \;\;\;\;\;\;
\alpha'_C = \alpha_Z Q_{c}  \;\;\;\;\;\;
\alpha'_\epsilon=  \alpha_\epsilon \xi_{\epsilon}
\end{eqnarray}
and are shown in Fig.\ref{TransportEnergyFinal}.


The integration conditions at the center of the sun ($r=0$) are given by:
\begin{eqnarray}
\nonumber
\delta m &=& \gamma_{P,0}\, \delta P_{0} + \gamma_{T,0} \, \delta T_{0}
+ \gamma_{Y,0} \, \Delta Y_{\rm ini}  + \gamma_{\epsilon,0} \; \delta \epsilon_{0} \\
\nonumber
\delta P &=& \delta P_{0}\\
\nonumber
\delta T &=& \delta T_{0}\\
\delta l &=& \beta'_{P,0} \,\delta P_{0} +  \beta'_{T,0} \,\delta T_{0} +
 \beta'_{Y,0} \,\Delta Y_{\rm ini} + \beta'_{C,0} \, \delta C + \beta'_{\epsilon,0} \,\delta \epsilon_{0}
\end{eqnarray}
where $\gamma_{P,0}=1.167$, $\gamma_{T,0}=-0.254$, $\gamma_{Y,0}=0.503$, $\gamma_{\epsilon,0}=0.161$ and
$\beta'_{P,0}=1.647$, $\beta'_{T,0}=1.884$, $\beta'_{C,0}=0.006$, $\beta'_{Y,0}=-1.142$, $\beta'_{\epsilon,0}=0.624$.
By imposing these conditions, the solution is obtained as a linear function of the four parameters $\Delta Y_{\rm ini}$, $\delta C$,
 $\delta T_{\rm 0}$, $\delta P_{\rm 0}$ which are fixed by requiring that:
\begin{eqnarray}
\nonumber
\delta m &=& - \overline{m}_{\rm conv} \, \delta C\\
\nonumber
\delta P  &=&  \delta C\\
\nonumber
\delta T &=& A'_Y \, \Delta Y_{\rm ini} +  A'_{C} \, \delta C\\
\delta l &=& 0
\end{eqnarray}
at the convective boundary ($r=\overline{R}_{\rm b}$), where $A'_{\rm Y} = -P_{Y,\rm b} \, A_{Y} = 0.626$,
$A'_{\rm C}=-P_{Y,\rm b} \, A_{\rm C} = 0.025$ and $\overline{m}_{\rm conv} =0.0192$.
The solution is, thus, univocally determined.

The chemical composition of the radiative region can be calculated by using rels.(\ref{helium}) and (\ref{MetalRad}). 
The variation of the density profile $\delta \rho(r)$ in the radiative region
can be obtained by the relation:
\begin{equation}
\label{densityFinal}
\delta \rho(r) = \gamma_P(r) \delta P(r) + \gamma_T(r) \delta T(r) + \gamma_Y(r) \Delta Y_{\rm ini} + \gamma_\epsilon(r) \delta \epsilon(r) \, .
\end{equation}
In the lower layers of the convective envelope the quantities $\delta P(r)$, $\delta T(r)$, $\delta \rho(r)$
are nearly constant (see sect.\ref{Boundary}),
while the quantity $\delta m(r)$ gradually goes to zero, as it is expected being the total mass of the
sun observationally determined. The surface chemical composition
can be determined by using rel.(\ref{ChimSurfFinal}).
We recall that the variations of surface helium and metal abundances are related among each other, 
in such a way that the hydrogen-to-metal surface ratio is unchanged with respect to the SSM.

Finally, the variation of the convective radius is calculated by using rel.(\ref{ConvRadius}). We obtain:
\begin{equation}
\label{ConvRadiusFinal}
\delta R_{\rm b} = \Gamma_{Y}\, \Delta Y_{\rm ini} + \Gamma_{C} \, \delta C + \Gamma_{\kappa}\, \delta \kappa_{\rm b}
\end{equation}
where
\begin{eqnarray}
\nonumber
\Gamma_{Y} & = & -\frac{A_Y}{\zeta_{\rm b}}\left[-P_{Y,\rm b} \, (\kappa_{T,\rm b}-4) + \kappa_{Y,\rm b} -\frac{\kappa_{Z,\rm b}}{1-\overline{Y}_{\rm b}}\right] = 0.449\\
\nonumber
\Gamma_{C} & = & -\frac{A_{C}}{\zeta_{\rm b}} \left[-P_{Y,\rm b} \, (\kappa_{T,\rm b}-4) + \kappa_{Y,\rm b} -\frac{\kappa_{Z,\rm b}}{1-\overline{Y}_{\rm b}}\right] -\frac{\kappa_{\rho,\rm b}+1}{\zeta_{\rm b}} = -0.117\\
\Gamma_{\kappa} & = & -\frac{1}{\zeta_{\rm b}} = -0.085
\end{eqnarray}
In the derivation of the above relation, we considered that $\delta l_{\rm b}=0$, $\delta m_{\rm b}\simeq 0$ and we used eqs.(\ref{surface}) and (\ref{ChimSurfFinal}).

\section{Comparison with full non linear calculations}
\label{Comparison}

In oder to show the validity of the proposed approach, we consider four possible modifications of the
standard input and we compare the results obtained by LSMs with those obtained by the full
non linear solar model calculations.
Three of the considered cases concern with opacity modifications that
are generally described as:
\begin{equation}
\kappa(\rho,T,X_{i})=F(T) \,\overline{\kappa}(\rho,T,X_{i})
\end{equation}
where $\overline{\kappa}(\rho,T,X_{i})$ represent the standard value and $F(T)$ is a
suitable function of the temperature. Namely, we consider:\\

\noindent
{\bf OPA1} - Overall rescaling of the opacity by a constant factor $F(T)\equiv 1.1$.
 This clearly corresponds to introduce the opacity variation:
$$
\delta \kappa(r)\equiv 0.1
$$
in eqs.(\ref{linsyst3}) for LSM calculations.

\begin{figure}[t]
\par
\begin{center}
\includegraphics[width=8.5cm,angle=0]{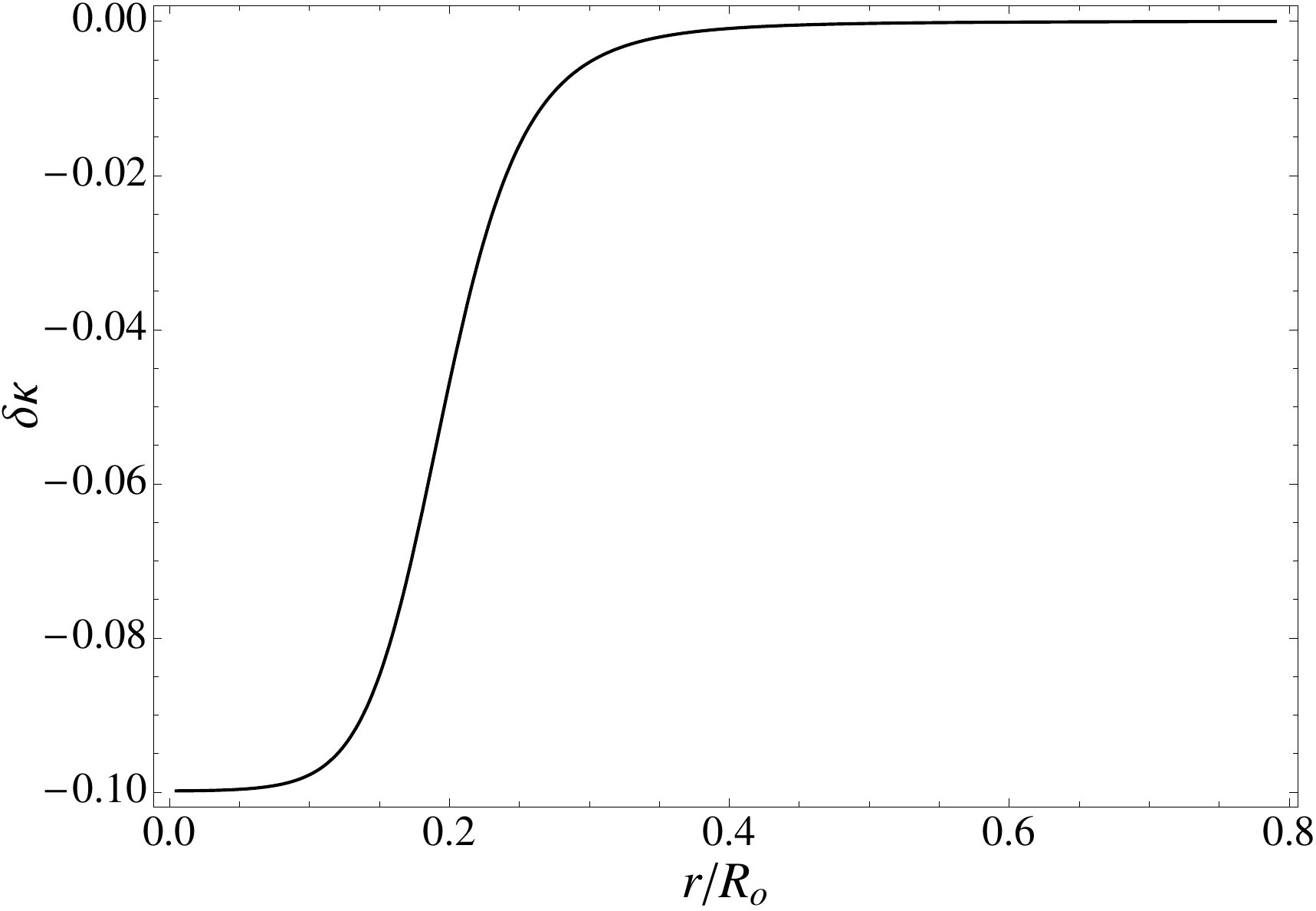}
\includegraphics[width=8.5cm,angle=0]{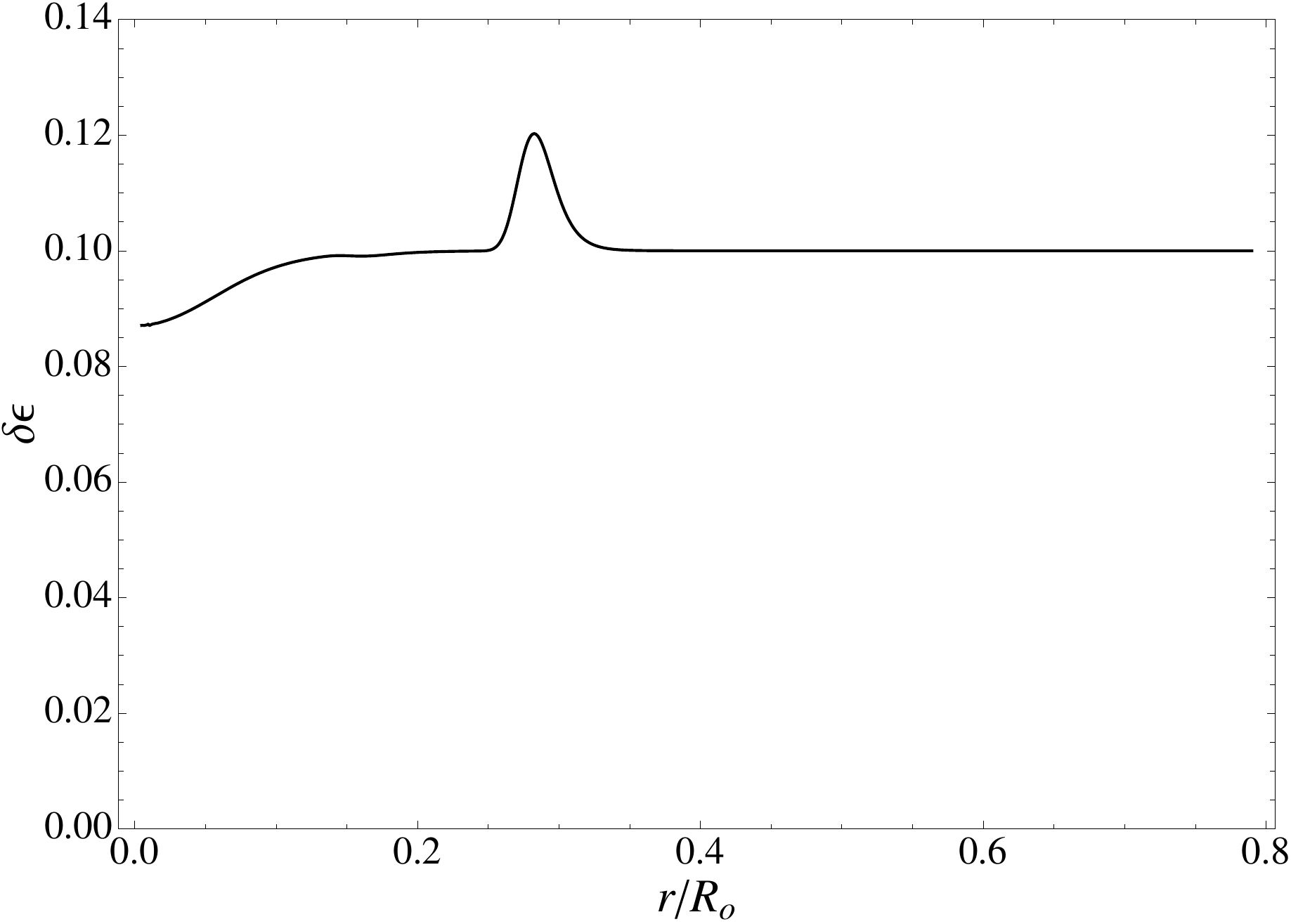}
\end{center}
\par
\vspace{-5mm} \caption{\em {\protect\small Left panel: The variation of the opacity profile $\delta \kappa(r)$ corresponding to the {\rm OPA2} case.
Right Panel: The variation energy generation profile $\delta \epsilon(r)$ corresponding to the {\rm Spp} case.}}
\label{OPA2andSPP}
\end{figure}

\noindent
{\bf OPA2} - Smooth decrease of the opacity at the center of the sun described by the function:
$$
F(T) = 1 + \frac{A}{1+\exp{\left[\frac{T_{\rm c} -  T} {fT_{\rm c}}\right]}}
$$
where $A=-0.1$, $T_{\rm c}=9.4\times 10^6$ K, $f=0.1$. In our approach this correspond to a
variation $\delta k(r)$  along the solar profile given by:
$$
\delta \kappa(r) = \frac{A}{1+\exp{\left[\frac{T_{\rm c} - \overline{T}(r)} {fT_{\rm c}}\right]}}
$$
where $\overline{T}(r)$ is the temperature profile predicted by the SSM (see left panel of Fig.\ref{OPA2andSPP}).

\noindent
{\bf OPA3} - Sharp increase of the opacity at the bottom of the convective envelope described
by assuming $F(T)=1.1$ for $T \le  5\times 10^{6} \, {\rm K}$ (and $F(T) =1$ otherwise).
 In our approach, this corresponds to:
\begin{eqnarray}
\nonumber
\delta \kappa(r) &=& 0.1 \;\;\;\;\;\;\; {\rm if} \;\;\;\;  \overline{T}(r) \le  5\times 10^{6}\, {\rm K} \\
\nonumber
\delta \kappa(r) &=& 0  \;\;\;\;\;\;\;\;\;\;\; {\rm otherwise}
\end{eqnarray}
The above inequality can be rewritten in terms of a condition on the distance from the center of the sun
obtaining $\delta \kappa(r) = 0.1$ for $r \ge 0.4 R_{\odot} $ (and $\delta \kappa(r)=0$ otherwise).\\


The fourth studied case concerns with modification of energy generation in the sun. More precisely, we consider:

\noindent
{\bf Spp} - Increase of the astrophysical factor $S_{\rm pp}$ of the $p+p\rightarrow d +e^+ + \nu_e$ reaction by +10\%.
In order to introduce this effect in LSM calculations, we consider the following variation of the energy generation profile:
\begin{equation}
\delta \epsilon(r) =\epsilon_{S_{\rm pp}}(r) \;\delta S_{\rm pp}
\end{equation}
where
\begin{equation}
\epsilon_{S_{\rm pp}}(r) = \frac{\partial \ln \epsilon}{\partial \ln S_{\rm pp}} |_{\rm SSM}
\end{equation}
and $\delta S_{\rm pp}=0.1$. The function $\delta \epsilon(r)$ obtained in this way is shown in the right
 panel of Fig.\ref{OPA2andSPP}. The bump at $r\simeq 0.28 R_{\odot}$ is due to the out-of-equilibrium behaviour of
helium-3.


\begin{figure}[t]
\par
\begin{center}
\includegraphics[width=8.5cm,angle=0]{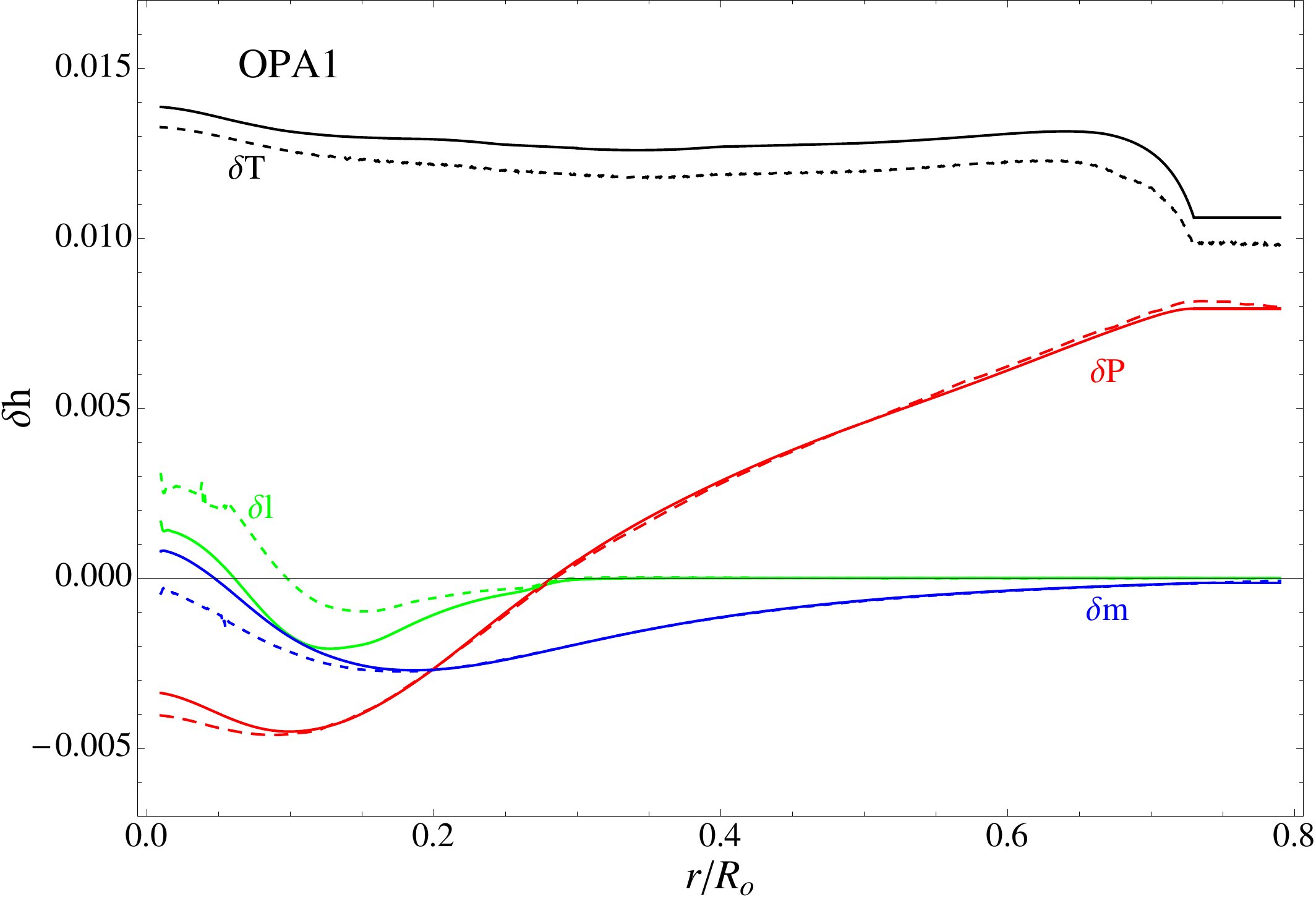}
\includegraphics[width=8.5cm,angle=0]{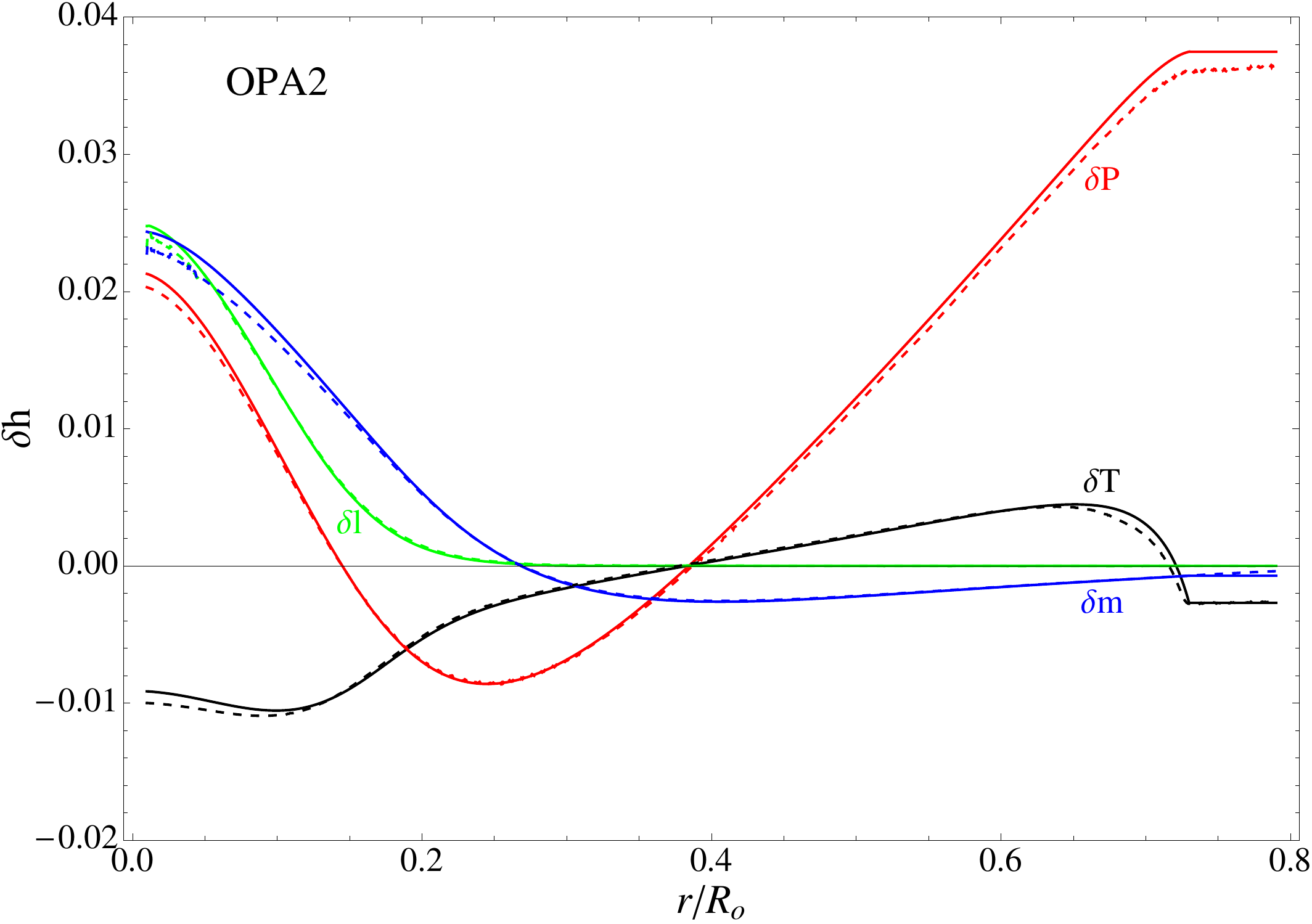}
\includegraphics[width=8.5cm,angle=0]{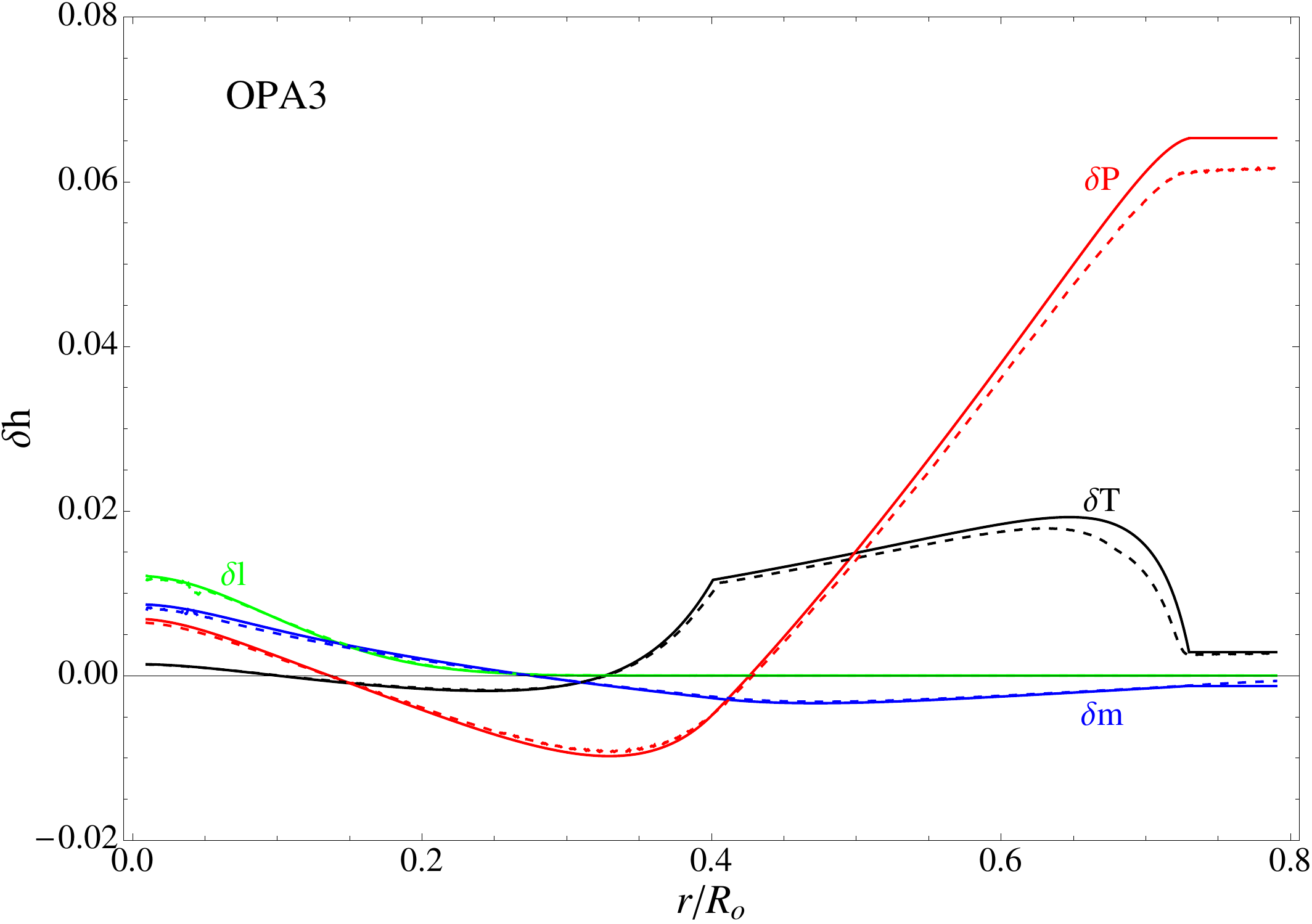}
\includegraphics[width=8.5cm,angle=0]{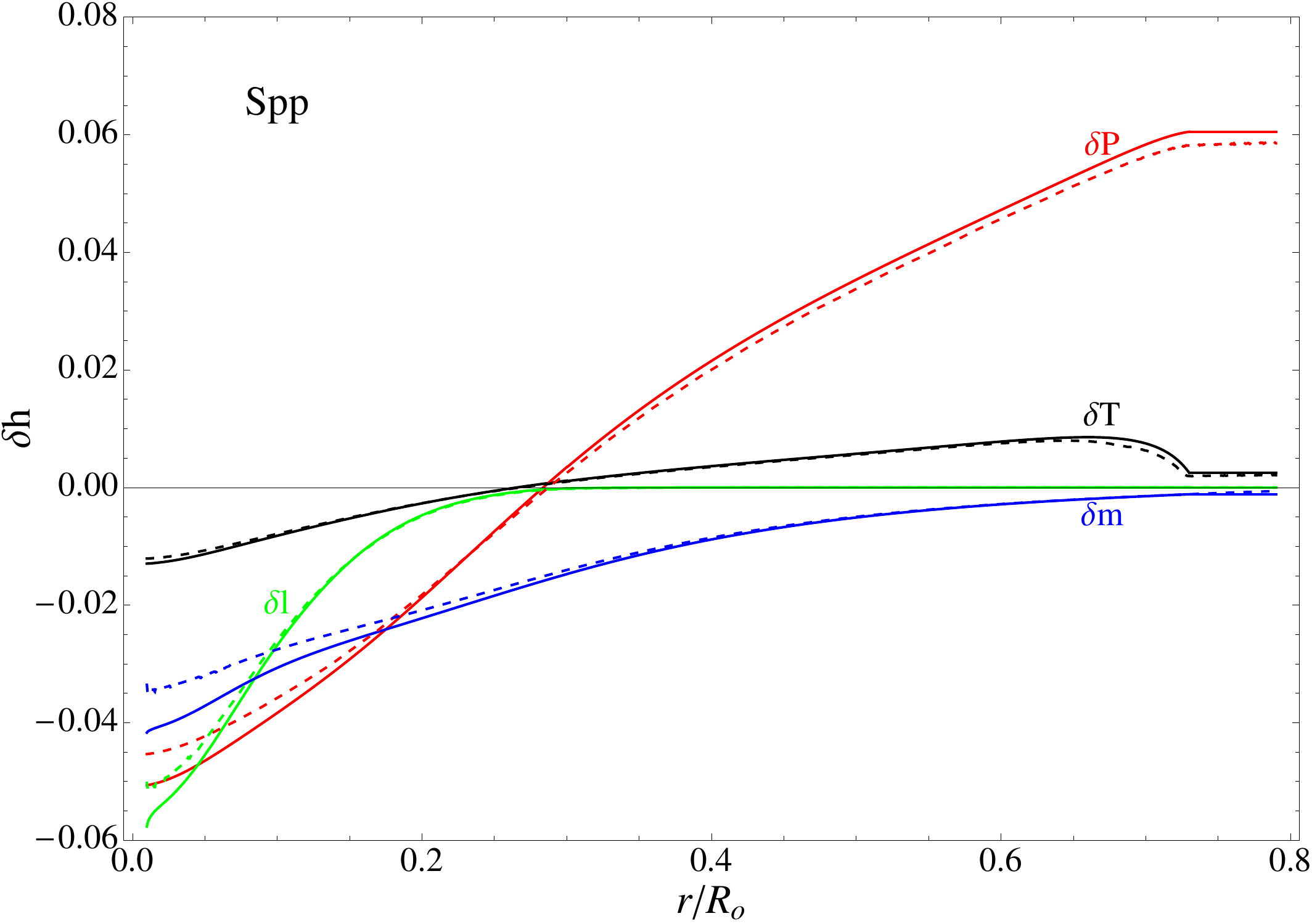}
\end{center}
\par
\vspace{-5mm} \caption{\em {\protect\small Comparison between the physical properties of the sun predicted by LSM (solid lines)
and by ``standard'' non-linear SM (dashed lines). In different colors we show the variations of pressure (red), temperature (black), mass (blue) and
luminosity (green) as a function of the solar radius.}}
\label{FigPhys}
\end{figure}



\begin{figure}[t]
\par
\begin{center}
\includegraphics[width=8.5cm,angle=0]{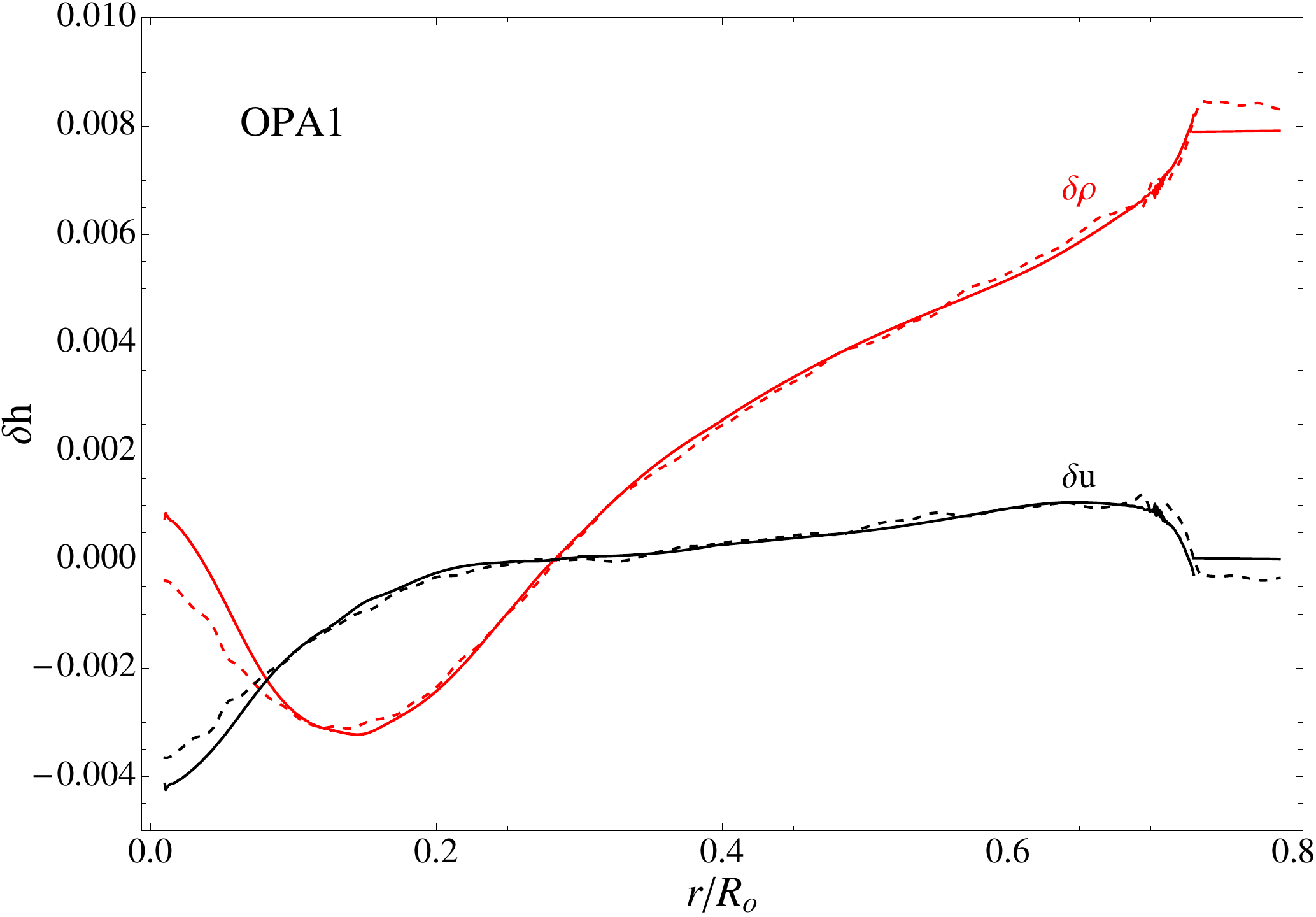}
\includegraphics[width=8.5cm,angle=0]{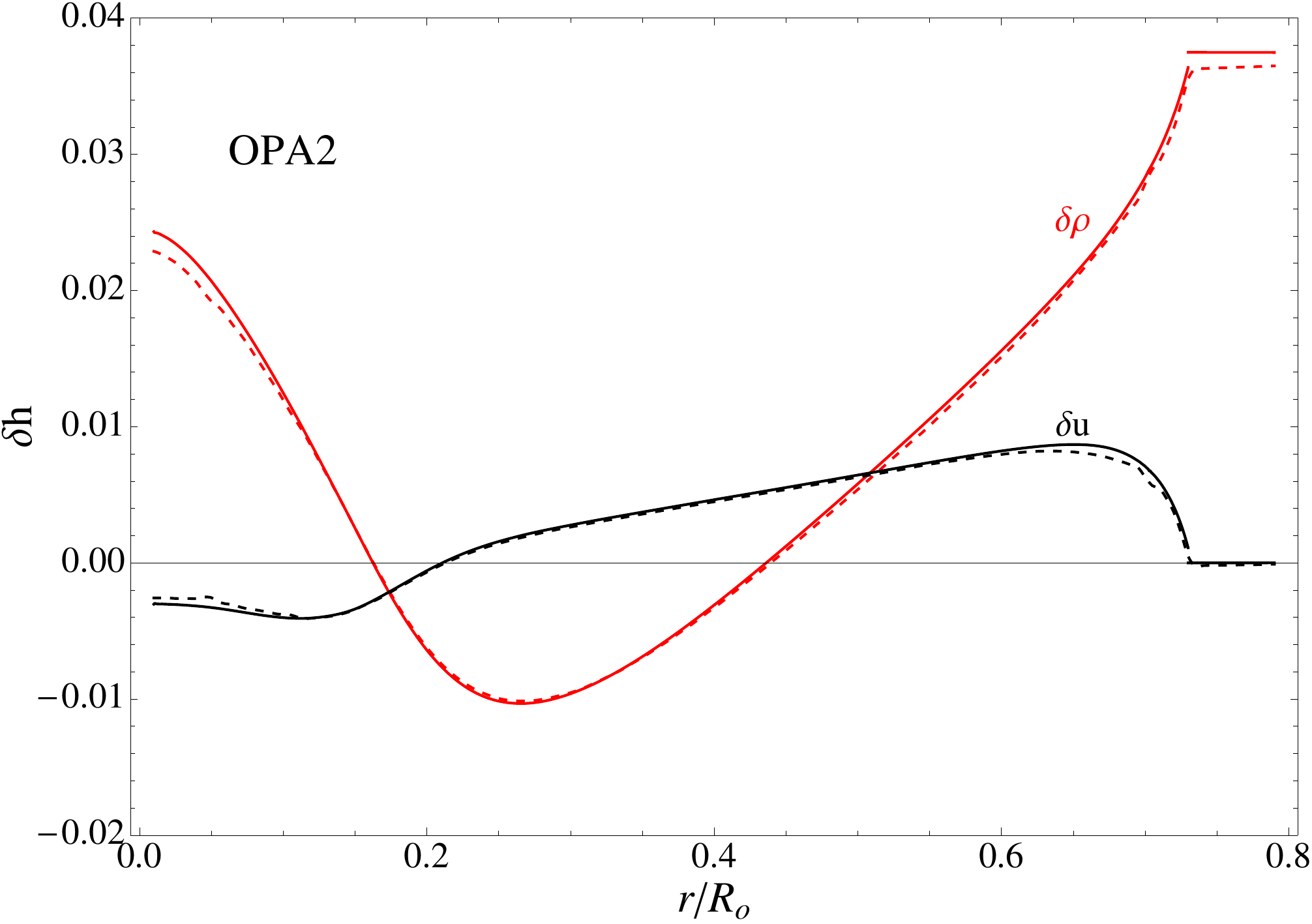}
\includegraphics[width=8.5cm,angle=0]{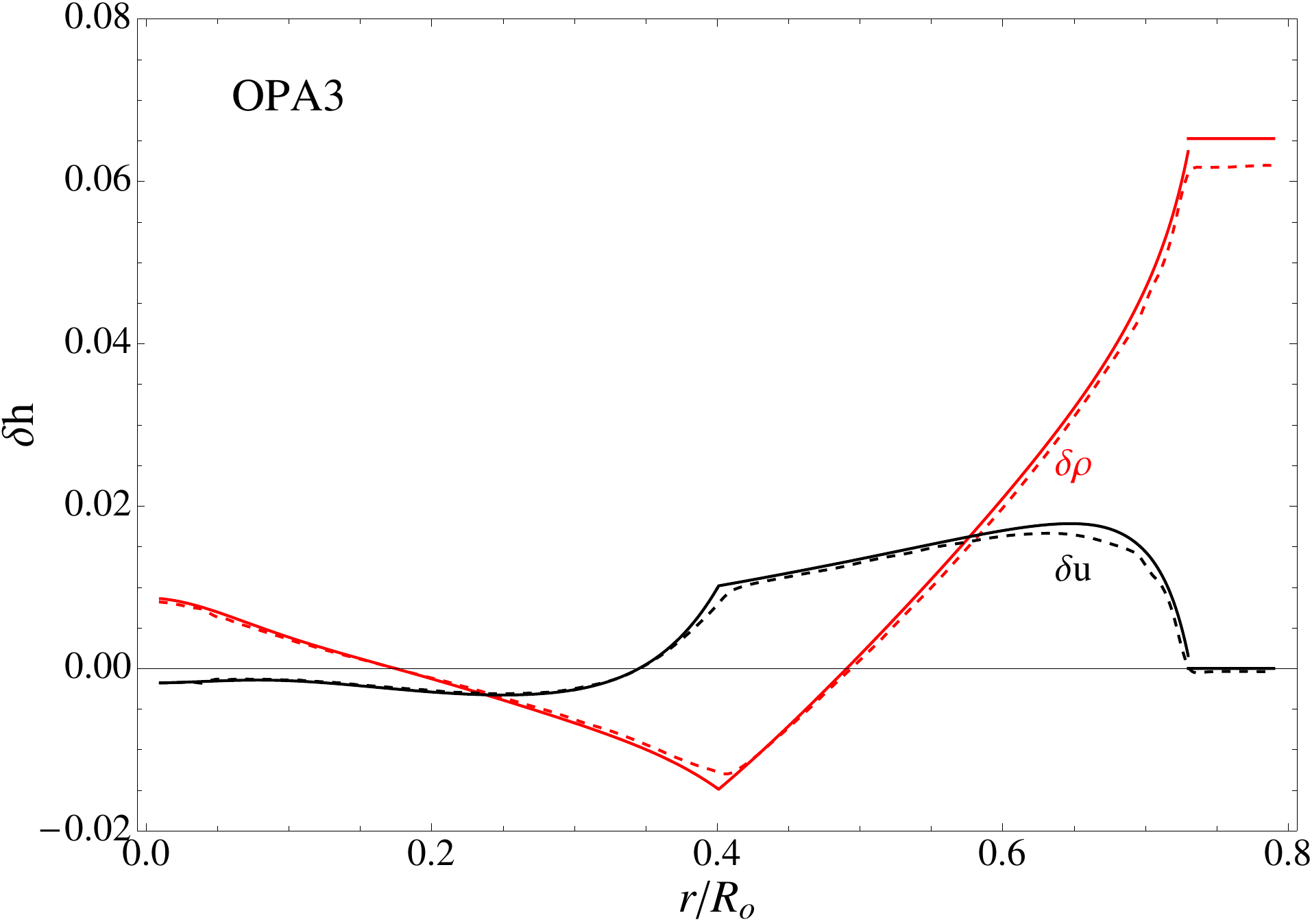}
\includegraphics[width=8.5cm,angle=0]{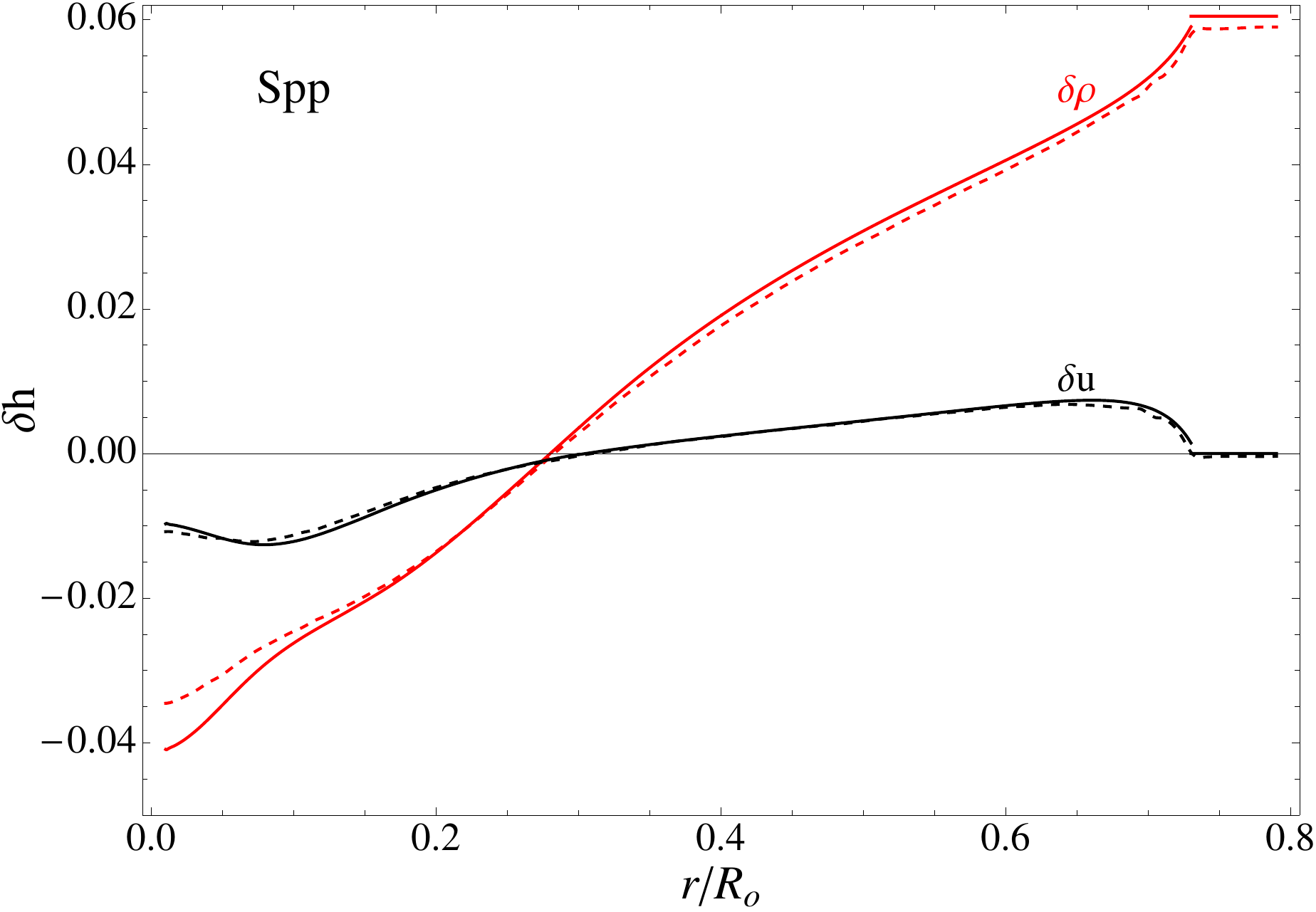}
\end{center}
\par
\vspace{-5mm} \caption{\em {\protect\small Comparison between the variations of density (red) and of squared isothermal sound speed (black)
predicted by LSM (solid lines)
and by ``standard'' non-linear SM (dashed lines).}}
\label{FigU}
\end{figure}


\subsection{Physical and helioseismic properties of the sun}

In Fig.\ref{FigPhys} we show with solid lines the physical properties of LSMs,
obtained by solving the linear system of eqs.(\ref{linsyst3}), and with dotted lines
the results obtained by the ``standard'' non-linear solar model (SM) calculations\footnote{The term ``standard'' is 
used here to refer to the conventional way of calculating solar models, see sect.\ref{SSM}. 
For the sake of precision, we use the acronym ``SM'' to refer to these models. 
This is done to avoid confusion with the  Standard Solar Model (SSM) which is intended as 
our best possible model of the sun, i.e. the model calculated in the conventional way and, moreover, 
by using the best possible choice for the input parameters.}.
The four panels corresponds to the input modifications introduced in the previous section.
We use different colors to show the variations of pressure (red), temperature (black), mass (blue) and
luminosity (green) as a function of the solar radius.

 We see that a very good agreement exists between LSM and non-linear SM results.
To be more quantitative, the response of the sun to the input
modifications is reproduced at the 10\% level or better, in all the considered cases.
It is important to note that all the relevant features of the $\delta h(r)$ are obtained, indicating that
the major effects are correctly implemented in our approach. The small differences between LSMs and 
non-linear SMs, typically more evident just below the convective region and/or at the center of the sun,
basically reflects the accuracy of the assumptions which have been used to estimate the variation of
the chemical composition of the present sun. 
In the OPA1 case, a disagreement exists in the $\delta l(r)$ behaviour at the center of the sun.
This difference, which has no observable consequences, is mainly due to non linear effects. 
One should note, in this respect, that the constant $10\%$ increase of the radiative opacity induces variations of temperature
and pressure which are at the $\sim 1\%$ level and a much smaller variation of $\delta l(r)$ (at the 0.1\% level). 
In this situation, cancellations between different  first-order competing contributions may occur.

\begin{table}[t]
\begin{center}{
\begin{tabular}{l|cc|cc|cc|cc}
& \multicolumn{2}{c}{OPA1}
& \multicolumn{2}{c}{OPA2}
& \multicolumn{2}{c}{OPA3}
& \multicolumn{2}{c}{Spp} \\
&  SM & LSM
&  SM & LSM
&  SM & LSM
&  SM & LSM\\
\hline
$\Delta Y_{\rm ini}$
&  0.016 & 0.017
& -0.0056 & -0.0058
& 0.0021 & 0.0019
&  0.0015& 0.0016 \\
$\delta Z_{\rm ini}$
& -0.018& -0.016
& 0.000 & -0.001
&-0.012&-0.012
&  -0.010&-0.011 \\
\hline
 $\Delta Y_{\rm b}$
 &  0.014 & 0.014
 & -0.0037 & -0.0036
 &0.0038 & 0.0038
& 0.0031 & 0.0034\\
$\delta Z_{\rm b}$
&  -0.018 & -0.018
&    0.0049 &   0.0047
&-0.0050 & -0.0049
&  -0.0040 & -0.0044   \\
$\delta R_{\rm b}$
& -0.0020  & -0.0020
& -0.0067  & -0.0070
& -0.014    & -0.015
& -0.0058  & -0.0064 \\
\hline
$\delta \Phi_{\rm pp}$
&-0.011 & -0.010
& 0.0045  & 0.0052
& -0.0020 & -0.0011
&  0.0090 & 0.0092\\
$\delta \Phi_{\rm Be}$
&0.13  & 0.13
&-0.067  & -0.064
&0.017  & 0.016
&-0.11 & -0.11\\
$\delta \Phi_{\rm B}$
&0.27  & 0.27
&-0.17  & -0.17
& 0.029 & 0.028
& -0.27 & -0.28\\
$\delta \Phi_{\rm N}$
&0.14  & 0.14
&-0.10  & -0.094
& 0.003 & 0.004
& -0.21& -0.22\\
$\delta \Phi_{\rm O}$
&0.21  & 0.22
& -0.14 & -0.14
& 0.012 & 0.012
& -0.29 & -0.31 \\
\hline
\end{tabular}
}\end{center}\vspace{0.4cm} \caption{\em {\protect\small
\label{suntab} Comparison between the predictions of LSM and ``standard'' non-linear SM for
the initial and surface chemical abundances, the convective radius
and the solar neutrino fluxes.
Note that the absolute variations are reported for Helium, whereas
the relative variations are shown for all the other quantities.
}}\vspace{0.4cm}
\end{table}

In Tab.\ref{suntab}, we present the variations of the initial abundances and of the photospheric
chemical composition obtained by using rels.(\ref{MetalRad}) and (\ref{ChimSurfFinal}).
Moreover, we show the variation of the convective radius $\delta R_{\rm b}$ calculated according to eq.(\ref{ConvRadiusFinal}).
Finally, in Fig.\ref{FigU} we present the variations of the density profile $\delta \rho(r)$
calculated according to eq.(\ref{densityFinal}) and the variation of the squared isothermal sound speed $u=P/\rho$
given by:
\begin{equation}
\delta u(r) = \delta P(r) - \delta \rho(r) \, .
\end{equation}
 We see that LSMs reproduces the results obtained by ``standard'' calculations with a very
good accuracy. In particular, the variations of the helioseismic observables are
obtained within $10\%$ or better (unless they are extremely small).
All this shows that our approach is sufficiently accurate to use LSMs as tool to investigate
the origin of the present discrepancy with helioseismic data. In a separated paper,
we will use LSMs to analyze the role of opacity and metals in the sun \cita{noiNext}.


\begin{figure}[t]
\par
\begin{center}
\includegraphics[width=8.5cm,angle=0]{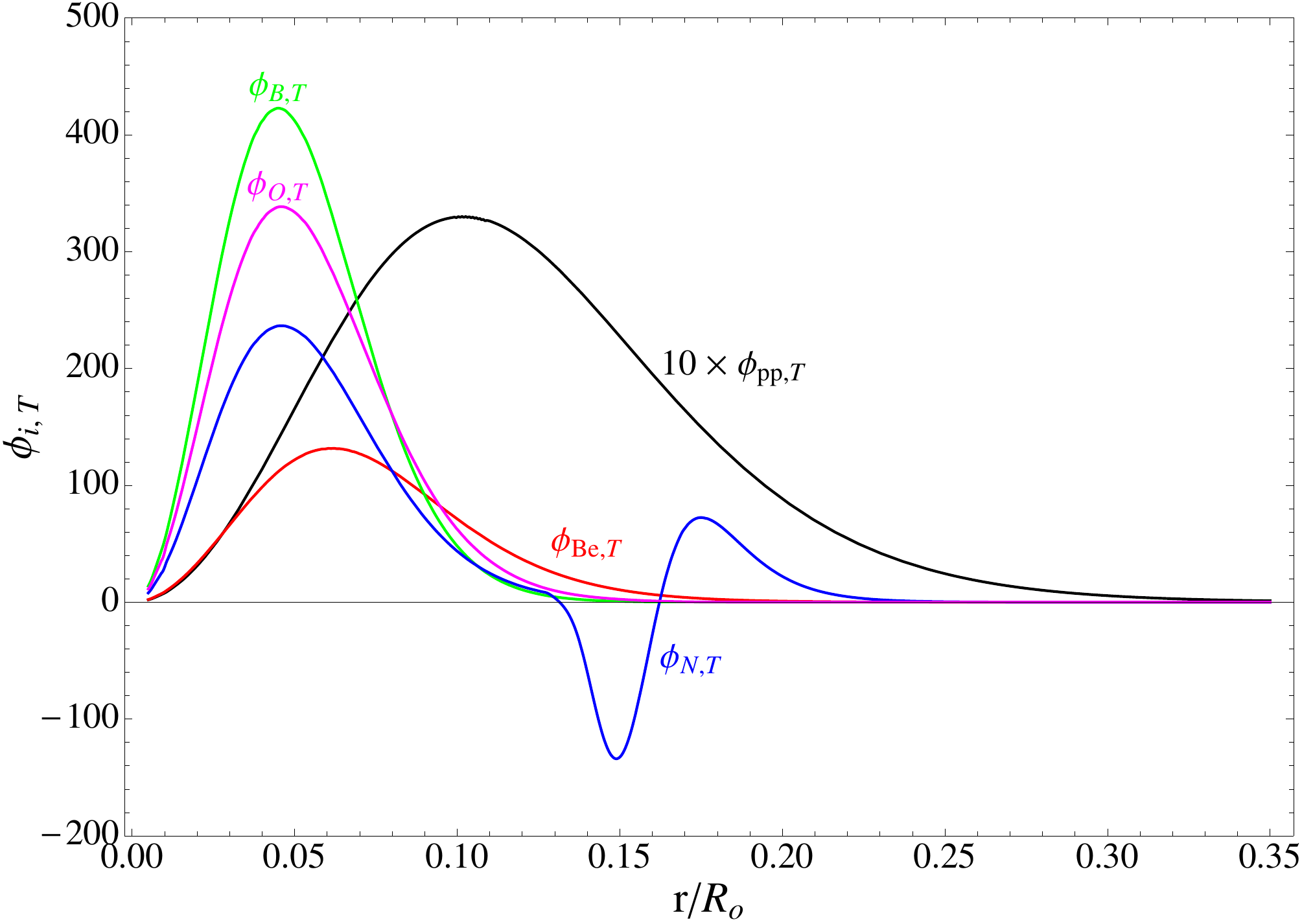}
\includegraphics[width=8.5cm,angle=0]{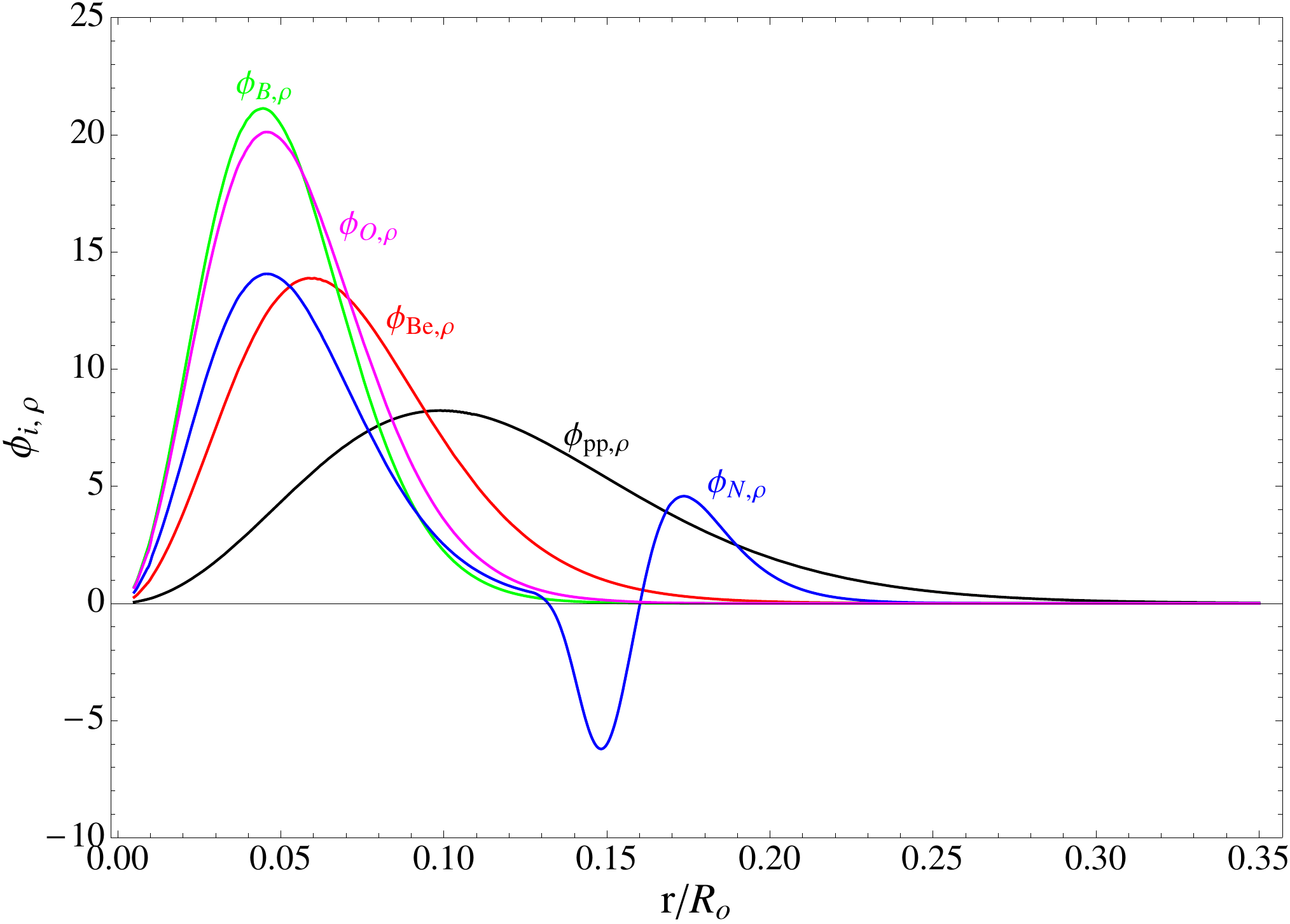}
\includegraphics[width=8.5cm,angle=0]{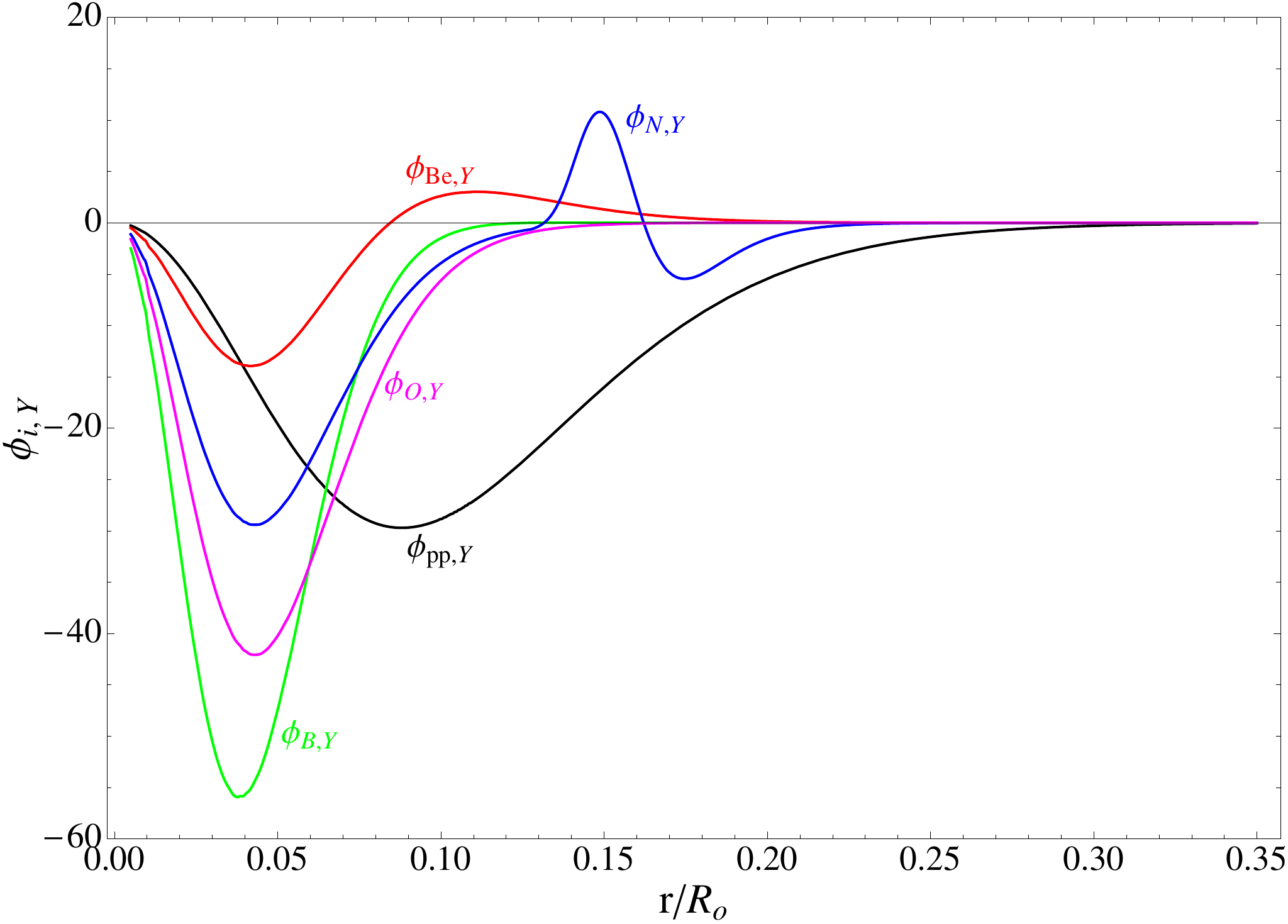}
\includegraphics[width=8.5cm,angle=0]{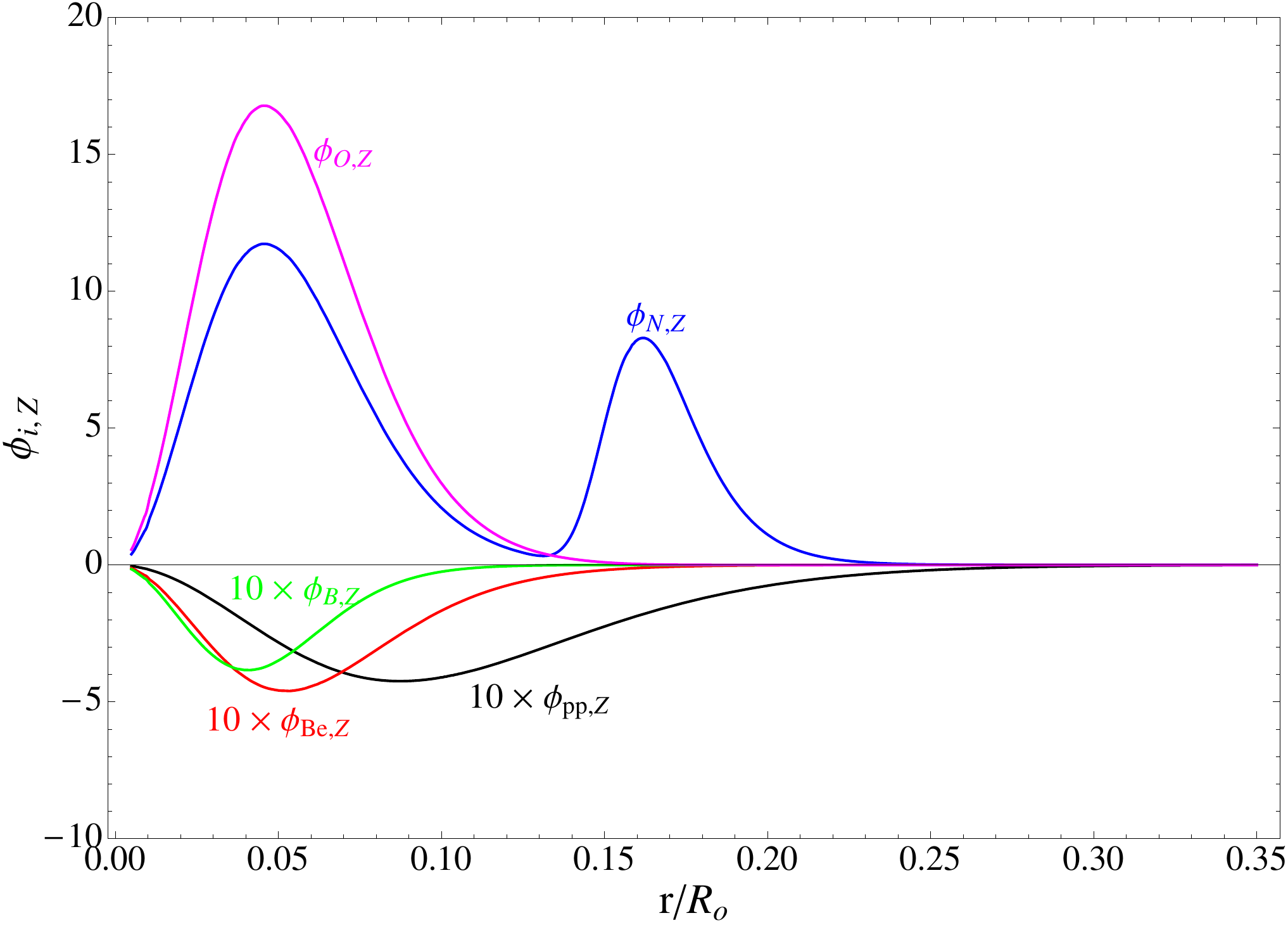}
\end{center}
\par
\vspace{-5mm} \caption{\em {\protect\small The functions $\phi_{\nu,j}(r)$ defined by eq.(\ref{phinuj}).}}
\label{Fignu}
\end{figure}


\subsection{Neutrino Fluxes}
\label{Neutrino}

As a final application, we calculate the solar neutrino fluxes predicted by LSMs and
we compare them with those obtained by using ``standard'' non-linear SM calculations.
The flux $\Phi_\nu$  of solar neutrinos can be expressed as:
\begin{equation}
\Phi_\nu = \frac{1}{D^2} \int dr \; r^2 \rho(r) \, n_{\nu}(r)
\end{equation}
where $D$ is the sun-earth distance, $n_{\nu}(r)$
is the total number of neutrinos produced per unit time and unit mass in the sun
and the index $\nu = {\rm pp, B, Be, N, O}$ labels the neutrino producing reaction
according to commonly adopted notations.
If we expand to first order, the relative variation of the
flux $\delta \Phi_\nu$  can be expressed as:
\begin{eqnarray}
\nonumber
\delta \Phi_\nu &=& \frac{1}{\overline{\Phi}_{\nu}}
\int dr \; (r/D)^2  \,  \overline{\rho}(r)\, \overline{n}_{\nu}(r) \times \\
& & \left[ n_{\nu,\rho}(r) \, \delta \rho(r)  + n_{\nu, T}(r)\,\delta T(r)
n_{\nu,Y}(r)\,\Delta Y(r) + n_{\nu,Z}(r) \,\delta Z(r) + n_{\nu,\rm Spp}(r)\, \delta S_{\rm pp}  \right]
\label{neutrino1}
\end{eqnarray}
where $\overline{\Phi}_{\nu}$ is the SSM value and:
\begin{eqnarray}
\nonumber
n_{\nu, \rm \rho}(r)&=&\left.\frac{\partial \ln n_\nu}{\partial \ln \rho}\right|_{\rm SSM}+1\\
\nonumber
n_{\nu,\rm T}(r)&=&\left.\frac{\partial \ln n_\nu}{\partial \ln T}\right|_{\rm SSM}\\
\nonumber
n_{\nu,Z}(r)&=&\left.\frac{\partial \ln n_\nu}{\partial \ln Z}\right|_{\rm SSM}\\
\nonumber
n_{\nu,Y}(r)&=&\left.\frac{\partial \ln n_\nu}{\partial Y}\right|_{\rm SSM}\\
\nonumber
n_{\nu,\rm Spp}(r)&=&\left.\frac{\partial \ln n_\nu}{\partial \ln S_{\rm pp}}\right|_{\rm SSM}\\
\end{eqnarray}
The derivatives $n_{\nu, j}(r)$ have been calculated numerically by assuming that the
abundances of secondary elements (helium-3, carbon-12 and nitrogen-14) 
can be estimated as it is described in appendix A.
The symbol $|_{\rm SSM}$ indicates that we calculate the derivatives 
along the density, temperature and
chemical composition profiles predicted by the SSM.

The term $n_{\nu,\rm Spp}(r) \, \delta S_{\rm pp}$ is introduced to describe the effects 
of a variation of $S_{\rm pp}$ on the rates of 
neutrino producing reactions. 
It clearly holds $n_{\rm pp,\rm Spp}(r) \equiv 1$,
since the ${\rm pp-}$neutrinos are produced by $p+p\rightarrow d + e^+ + \nu_e$. 
Boron and Beryllium neutrinos are influenced by $S_{\rm pp}$, because this parameter
determines the helium-3 production rate (through deuterium) and, thus, also the rate of
$^{3}{\rm He}+^4{\rm He}\rightarrow ^7{\rm Be}+\gamma$ reaction. As a consequence, 
we have:
\begin{equation}
n_{\rm Be, \rm Spp}(r) = n_{\rm B, \rm Spp}(r) =  \left.\frac{\partial \ln X_3}{\partial \ln S_{\rm pp}}\right|_{\rm SSM}
\end{equation}
where $X_{3}$ is the helium-3 abundance. The CN-cycle efficiency, instead,
does not depend on the value of $S_{\rm pp}$ and, thus, we have
$n_{\rm N,\rm Spp}(r) = n_{\rm O,\rm pp}(r) \equiv 0$.

It is useful to recast eq.(\ref{neutrino1}) in the form:
\begin{equation}
\delta \Phi_\nu =
\int dr \; \left[ \phi_{\nu,\rho}(r) \, \delta \rho(r) + \phi_{\nu, T}(r) \,\delta T(r) + \phi_{\nu,Y}(r) \,\Delta Y(r) + \phi_{\nu,Z}(r)  \,\delta Z(r) + \phi_{\nu,\rm Spp}(r) \, \delta S_{\rm pp}  \right]
\label{nuflux}
\end{equation}
where:
\begin{equation}
\phi_{\nu,j} (r) = \frac{r^2 \,  \overline{\rho}(r) \, \overline{n}_{\nu}(r) \, n_{\nu,j}(r)}{\int dr \; r^2 \,  \overline{\rho}(r) \, \overline{n}_{\nu}(r)}
\label{phinuj}
\end{equation}
The functions $\phi_{\nu,j}(r)$ are displayed in Fig.\ref{Fignu}. They show explicitly
the well-known fact that the boron, beryllium and CNO neutrinos are produced in a more internal region with respect to pp-neutrinos
and that they more strongly depend on temperature. The complicated behaviour of the functions $\Phi_{N, j}(r)$ at
$r\simeq 0.15 R_{\odot}$ is due to the out-of-equilibrium behaviour of carbon-12 abundance. 
The $N-$neutrinos originate, in fact, from the decay of $^{13}{\rm N}$, which is produced by $^{12}{\rm C}+p\rightarrow ^{13}{\rm N}+\gamma$
reaction. The sun was born with a relatively large amount of carbon-12, which has been converted by CN-cycle 
into nitrogen-14 in the more internal regions. The $^{12}{\rm C}$ abundance is, thus, larger where CN-cycle 
is less effective and,  as a consequence, a non negligible production of $N-$neutrinos occurs at relatively large radii, 
where equilibrium conditions do not hold.

In Tab.\ref{suntab}, we compare the LSM results, i.e. the values $\delta \Phi_\nu$ obtained by applying eq.(\ref{nuflux})
to the solutions of eqs.(\ref{linsyst3}), with the results obtained from ``standard'' calculations. 
We see that an excellent agreement exists, for all the cases considered in this paper.\footnote{It is remarkable the fact
that all the neutrino fluxes are correctly reproduced, even in the cases in which some compensations occur. See e.g. the
OPA3 case, in which a very small value for $\delta \Phi_{N}$ flux is obtained
(with respect to the other fluxes), as a consequence of the peculiar behaviour of the $\Phi_{N,j}(r)$ functions.}
All this show that our approach can be used as a tool to investigate the dependence of the solar neutrino
fluxes on the input parameters adopted in SSM construction.

\section{Conclusions}
\label{conclusions}

We have proposed a new approach to study the properties of the sun which is based on the following points:\\
{\em i)} We have considered small variations of the physical and chemical properties of the sun with respect to Standard Solar Model
predictions and we have linearized the stellar equilibrium equations to relate them to the properties of the solar plasma (see sects.\ref{Linear},\ref{Properties},\ref{Boundary});  \\
{\em ii)} We have derived simple relations which allow to estimate the (variation of) the present solar composition
from the (variation of) the nuclear reaction rates and elemental diffusion efficiency in the present sun (see sect.\ref{chemical}).\\
As a final result, we have obtained a linear system of ordinary differential equations (see sect.\ref{Fullset}) which can be easily solved and that 
completely determine the physical and chemical properties  of the ``perturbed'' sun.

In order to show the validity of our approach, we have considered four possible modifications of the input parameters
(opacity and energy generation profiles) and we have compared the results of our Linear Solar Models 
with those obtained by ``standard'' methods  for solar model calculations (see sect.\ref{Comparison}).
A very good agreement is achieved for all the structural parameters (mass, luminosity, temperature, pressure, etc.)
and for all the helioseismic and solar neutrino observables.

We believe that our approach can complement the traditional method for solar model calculations, allowing to investigate in a more efficient and 
transparent way the role of the different parameters and assumptions. In particular, it could be useful to study the origin of the 
present discrepancy between SSM results and helioseismic data.

\section*{\sf  Acknowledgment}
\def\refname{\vskip-1cm}
\baselineskip=1.15em

We thank A. Formicola from LUNA collaboration for providing us useful information about nuclear cross sections and
G. Fiorentini for comments and discussions. We are extremely grateful to S. Degl'Innocenti and to the other members 
of the Astrophysics group of University of Pisa for very useful discussions, for providing us updated opacity tables, 
for reading the manuscript and for fruitful collaboration.

\newpage

\section*{A: The abundance of secondary elements}

Secondary elements are those elements which are both created and destroyed in a reaction chain or in a reaction cycle. 
The relevant secondary elements to understand energy and neutrino production in the sun are $^{3}{\rm He}$, 
$^{12}{\rm C}$ and $^{14}{\rm N}$.

\subsection*{A1: The $^3$He abundance}

The evolution of helium-3 abundance, $X_3$, is described by the equation\footnote{We assume that deuterium is at equilibrium, 
i.e. deuterium production and destruction rates are equal.}:
\begin{equation}
\frac{\partial X_{3}}{\partial t} = \frac{3 \rho}{ m_{u}}\left[\frac{X^2}{2}\langle \sigma v \rangle_{\rm pp} - \frac{X_{3}^2}{9}\langle \sigma v \rangle_{33}
  - \frac{X_{3} Y}{12}\langle \sigma v \rangle_{\rm 34}  \right]
\end{equation} 
where $\langle \sigma v \rangle_{\rm pp}$ is the reaction rate per particle pair of the $p+p\rightarrow d + e^{+} + \nu_e$ reaction,
while $\langle \sigma v \rangle_{33}$ and $\langle \sigma v \rangle_{34}$ refer to $^{3}{\rm He}+^{3}{\rm He}\rightarrow ^{4}{\rm He} + 2p $ and 
$^{3}{\rm He}+^{4}{\rm He} \rightarrow ^{7}{\rm Be}+\gamma$ reactions, respectively.
In the most internal region ($r\le 0.25 R_{\odot}$), the rates of these reactions are fast with 
respect to the sun evolutionary times and equilibrium is achieved:
\begin{equation}
X_{3,\rm eq} = \frac{3 \, Y}{8} \frac{\langle \sigma v \rangle_{34} }{\langle \sigma v \rangle_{33} }\left[-1 + \sqrt{ 1+32 \left(\frac{X}{Y}\right)^2 \frac{\langle \sigma v \rangle_{33}\; \langle \sigma v \rangle_{\rm pp}} {\langle \sigma v \rangle_{\rm 34} ^2}}    \right]
\label{He3_eq}
\end{equation} 
In the outer regions ($R\ge 0.2 R_{\odot}$), we can neglect the contribution from 
$^{3}{\rm He}+^{4}{\rm He} \rightarrow ^{7}{\rm Be}+\gamma$ and we obtain:
\begin{equation}
\frac{\partial X_{3}}{\partial t} = {\mathcal C}_{3} - {\mathcal D}_{3} \; X_{3}^2 
\end{equation} 
where ${\mathcal C}_3 = ( 3\,  X^2 \, \rho \, \langle \sigma v \rangle_{\rm pp}  ) / ( 2\, m_{u} ) $ and ${\mathcal D}_3 =  (\rho \, \langle \sigma v \rangle_{\rm 33}  ) / ( 3\, m_{u} )$.
If we assume that the factors ${\mathcal D}_3$ e ${\mathcal C}_3$ are approximately constant 
over the time scale $\tau_3 = {\mathcal C_3}^{-1}$ during which $^3{\rm He}$ is produced, the above equation can be explicitly 
solved.  We obtain:
\begin{equation}
X_{3} = X_{3,\rm eq} \; \tanh \left[\frac{X_{3,\rm max} }{X_{3,\rm eq} }\right]
\label{He3_gen}
\end{equation}
where $X_{3,\rm max} = {\mathcal C_3} \, t$ represents the helium-3 abundance which would have been produced in the 
integration time $t$ if we had neglected the $^3{\rm He}$ destruction processes (and we implicitly assumed that the 
helium-3 initial abundance is negligible).

In this work, we need to calculate the helium-3 response to a generic modification of the solar properties.
We use the following approach.  We calculate the $^{3}{\rm He}$ equilibrium value in a 
generic solar model by using eq.(\ref{He3_eq}). 
We assume that eq.(\ref{He3_gen}) is valid at each point of the sun and we estimate 
the value $\overline{X}_{3,\rm max}$ in the SSM by inverting it 
(and by using the SSM-abundance $\overline{X}_3$).
Then, we calculate the value of $X_{3,\max}$ in the modified solar model by 
considering that $X_{3,\rm max}$ scales as $X_{3,\rm max}\propto   X^2 \, \rho \, S_{\rm pp}$. 
Finally, we calculate the value of $X_{3}$ in the modified sun by using eq.(\ref{He3_gen}) with the 
modified values for $X_{3,\rm eq}$ and $X_{3,\rm max}$. 
The obtained results reproduce with very good accuracy the variation of the $^{3}{\rm He}$ abundances in 
all the cases considered in this paper.

\subsection*{A2: The $^{12}$C and $^{14}$N abundances}

The evolution of carbon-12 and nitrogen-14 abundances, indicated with $X_{12}$ and $X_{14}$ respectively, is described by the 
equations\footnote{We neglect the NO-cycle, which is largely sub-dominant with respect to the CN-cycle in the sun.}:
\begin{eqnarray}
\nonumber
\frac{\partial X_{12}}{\partial t} &=& \frac{12 \, \rho}{ m_{u}}\left[\frac{X\,X_{14} }{14}\langle \sigma v \rangle_{1, 14} - \frac{X\, X_{12}}{12}\langle \sigma v \rangle_{1, 12}   \right]\\
\frac{\partial X_{14}}{\partial t} &=& \frac{14 \, \rho}{ m_{u}}\left[\frac{X\,X_{12} }{12}\langle \sigma v \rangle_{1, 12} - \frac{X\, X_{14}}{14}\langle \sigma v \rangle_{1, 14}   \right]
\end{eqnarray} 
where $\langle \sigma v \rangle_{1, 12}$ and $\langle \sigma v \rangle_{1, 14} $ are the reaction rates per particle pair of $^{12}{\rm C} + p \rightarrow ^{13}{\rm N} + \gamma$
and $^{14}{\rm N} + p \rightarrow ^{15}{\rm O} + \gamma$, respectively.
It is useful to rewrite these equations in terms of the variables $\eta = X_{14}/14 - X_{12}/12$ and $ N =  X_{12}/12 + X_{14}/14$. 
We obtain:
\begin{eqnarray}
\nonumber
\frac{\partial \eta}{\partial t} &=& {\mathcal C}_\eta \, N- {\mathcal D}_\eta \, \eta \\
\frac{\partial N}{\partial t} &=& 0
\label{eq_eta}
\end{eqnarray} 
where ${\mathcal C}_\eta = \rho \,  X (\langle \sigma v \rangle_{1, 12} - \langle \sigma v \rangle_{1, 14} ) / m_{u} $ and 
$ {\mathcal D}_\eta  = \rho \,  X (\langle \sigma v \rangle_{1, 14}  + \langle \sigma v \rangle_{1, 12} ) / m_{u}$. 
From the above equations, we see that $N$ is constant and that the equilibrium value for $\eta$ is given
by:
\begin{equation}
\eta_{\rm eq} = N\; \frac{\langle \sigma v \rangle_{1, 12} - \langle \sigma v \rangle_{1, 14} }{\langle \sigma v \rangle_{1, 12} + \langle \sigma v \rangle_{1, 14} }\; .
\label{eta_eq}
\end{equation}
In the assumption that the coefficients ${\mathcal C}_\eta$ and ${\mathcal D}_\eta$ are approximately constant 
over the time scale $\tau_{\eta} = 1/({\mathcal C}_\eta \, N)$, the solution for $\eta$ can be explicitly calculated, 
obtaining
\begin{equation}
\eta = \eta_{\rm eq} + (\eta_0 - \eta_{\rm eq}) \exp\left( - \frac{\eta_{\rm max}}{\eta_{\rm eq}}\right)
\label{eta_gen}
\end{equation}
where $\eta_0$ is the initial value and $\eta_{\rm max} = {\mathcal C}\, N\, t$ represents the $\eta$ value 
which would have been obtained in the integration time $t$ if we had neglected the 
``destruction'' term in eq.(\ref{eq_eta}). 

 In analogy to what was done for helium-3, we use the following approach 
to calculate the response of carbon-12 and nitrogen-14 to a generic modification of the solar properties.
We calculate the $\eta_{\rm eq}$ value in a generic solar model by using eq.(\ref{eta_eq}). 
We assume that eq.(\ref{eta_gen}) is valid at each point of the sun and we estimate the value 
$\overline{\eta}_{\rm max}$ in the SSM by inverting it (and using the SSM-values $\overline{\eta}$ and $\overline{\eta}_{0}$).
Then, we calculate the value of  $\eta_{\rm max}$ in the modified sun by considering that it 
scales as $\eta_{\rm max}\propto   N \, \rho \, X \, (\langle \sigma v \rangle_{1, 12}- \langle \sigma v \rangle_{1, 14})$. Finally,  
we calculate $\eta$ in the modified solar model by using eq.(\ref{eta_gen}) with the 
new values for $\eta_{\rm eq}$, $\eta_{\rm max}$ and $\eta_0$. 
The carbon-12 and nitrogen-14 abundances can be calculated from $\eta$ and 
$N$ by the simple relations:
\begin{eqnarray}
\nonumber
X_{12} &=& 6 (N - \eta)\\
X_{14} &=& 7 (N + \eta)
\end{eqnarray}
The obtained results reproduce with very good accuracy the results obtained by full numerical calculations
in all the cases considered in this paper.

\section*{B: Derivation of eq.(\ref{EQ_diffusion})}

The effect of elemental diffusion on the convective envelope chemical composition
can be estimated from:
\begin{equation}
\delta D_{i,\rm b} = \delta \omega_{i} - \delta H
\end{equation}
where $\delta H$ is the relative variation of the convective envelope effective thickness, while
$\delta \omega_{i}$ is the relative variation of the diffusion velocity of the $i-$element
at the bottom of the convective region.
One should note that the quantity $\delta \omega_{i}$ is defined as:
\begin{equation}
\delta \omega_{i} = \frac{\omega_i(R_{\rm b}) - \overline{\omega}_i(\overline{R}_{\rm b})}{\overline{\omega}_i(\overline{R}_{\rm b})}
\end{equation}
and, thus, involves the difference between the diffusion velocities 
evaluated at two different points. By taking into account the expression for $\omega_i$ given in eq.(\ref{omega}) and 
by considering that $\partial \ln P/ \partial r \simeq  - (G_{\rm N} M_{\odot} / R_{\rm b}^2) (\rho/P) $
at $r = R_{\rm b}$, we obtain:
\begin{equation}
\delta \omega_{i} = \frac{5}{2}\,\frac{ T(R_{\rm b} )-\overline{T}(\overline{R}_{\rm b}) }{\overline{T}(\overline{R}_{\rm b})} 
-\frac{ P(R_{\rm b} ) -\overline{P}(\overline{R}_{\rm b}) }{\overline{P}(\overline{R}_{\rm b})}  -2\, \delta R_{\rm b} + \frac{d \ln A_{\rm tot, i} }{d Y}\,\Delta Y_{\rm b}
\label{domega0}
\end{equation}
where $A_{\rm tot, i}= A_{P, i} + \nabla_{\rm ad} \, A_{T, i}$.
This can be rewritten as:
\begin{eqnarray}
\nonumber
\delta \omega_{i} &=& \frac{5}{2}\,\delta T_{\rm b} -\delta P_{\rm b} -2\, \delta R_{\rm b} + \frac{d \ln A_{\rm tot, i} }{d Y}\,\Delta Y_{\rm b} \\
                            &  & + \left[\frac{5}{2}\frac{\partial \ln \overline{T}(\overline{R}_{\rm b})}{\partial \ln r} - 
                                       \frac{\partial \ln \overline{P}(\overline{R}_{\rm b})}{\partial \ln r}  \right] \delta R_{\rm b}
\label{domega}
\end{eqnarray}
where, according to the notation convention adopted in this paper, we have indicated with 
$\delta T_{\rm b} = [ T(\overline{R}_{\rm b})- \overline{T}(\overline{R}_{\rm b}) ] / \overline{T}(\overline{R}_{\rm b})$
and $\delta P_{\rm b} = [ P(\overline{R}_{\rm b})- \overline{P}(\overline{R}_{\rm b}) ] / \overline{P}(\overline{R}_{\rm b})$.\footnote{We 
assumed that $\partial \ln T (R_{\rm b}) /\partial \ln r \simeq \partial \ln \overline{T}(\overline{R}_{\rm b}) /\partial \ln r $ 
and $\partial \ln P(R_{\rm b}) /\partial \ln r \simeq \partial \ln \overline{P}(\overline{R}_{\rm b}) /\partial \ln r $.}
We note that the last term in the above equation can be neglected since
$\partial \ln T / \partial \ln P = \nabla_{\rm ad}\simeq 2/5$ at the bottom of the convective region.

The effective thickness of the convective envelope is defined by the condition  
$H=M_{\rm conv}/(4\pi R_{\rm b}^2 \rho_{\rm b})$. The relative variation $\delta H$ is thus given by
\begin{equation}
\delta H = \delta M_{\rm conv} - 2 \delta R_{\rm b} - \delta \rho_{\rm b} -\frac{\partial \ln \rho (\overline{R}_{\rm b})}{\partial \ln r} \, \delta R_{\rm b}
\label{dH}
\end{equation}
If we consider that $M_{\rm conv}= \int_{R_{\rm b}} dr \; 4\pi r^2 \rho(r)$ and that 
$\delta \rho(r)$ is approximately constant in the convective region, we obtain:
\begin{equation}
\delta M_{\rm conv} = \delta \rho_{\rm b} - \frac{\overline{R}_{\rm b}}{\overline {H}}\,\delta R_{\rm b}
\end{equation}
By using this relation and taking into account eqs.(\ref{dH}) and (\ref{domega}), we arrive at the expression:
\begin{equation}
\delta D_{i,\rm b} = \frac{5}{2}\,\delta T_{\rm b} -\delta P_{\rm b} + \frac{d \ln A_{\rm tot, i}}{d Y}\, \Delta Y_{\rm b} + 
\left[\frac{\partial \ln \rho (\overline{R}_{\rm b})}{\partial \ln r} + \frac{\overline{R}_{\rm b}}{\overline {H}} \right] \delta R_{\rm b}
\end{equation}
Finally, by using the expression for $\delta R_{\rm b}$ given in eq.(\ref{ConvRadius}) (with $\delta m_{\rm b} = 0$ and $\delta l_{\rm b} = 0$) 
and by taking advantage of the integration conditions at the bottom of the convective region expressed 
by eqs.(\ref{surface}), we obtain the final relations:
\begin{eqnarray}
\nonumber
\delta D_{Y,\rm b} &=& \Gamma_{Y} \; \delta T_{\rm b} +  \Gamma_{P} \, \delta P_{\rm b}  + \Gamma_{\kappa}\, \delta \kappa_{\rm b} \\
\delta D_{Z,\rm b} &=& \Gamma_{Z} \; \delta T_{\rm b} +  \Gamma_{P} \, \delta P_{\rm b}  + \Gamma_{\kappa}\, \delta \kappa_{\rm b} 
\end{eqnarray}
where $\Gamma_{Y} = 2.05 $, $\Gamma_{Z}= 2.73$, $\Gamma_{P}= -1.10$ and $\Gamma_\kappa = -0.06$
and we assumed that all metals have the same diffusion velocity as iron.
In the calculations presented in this paper, we neglect for simplicity the terms proportional to
 $\delta \kappa_{\rm b}$ which generally give a very small contribution.

\newpage

\section*{\sf  References}
\def\refname{

\end{document}